\documentclass[review]{elsarticle}
\usepackage{graphicx}% Include figure files
\usepackage{dcolumn}% Align table columns on decimal point
\usepackage{bm}% bold mathf
\usepackage{braket}% bra-ket (Dirac) notation
\usepackage{graphicx}% Include figure files
\usepackage{epstopdf}
\usepackage{mwe}    % loads »blindtext« and »graphicx«
\usepackage{subfig}
\usepackage[T1]{fontenc}
\usepackage{caption}
\usepackage{amssymb}
\usepackage{amsmath}
\usepackage{booktabs}
\usepackage{tabularx}
\usepackage{multirow}

\newcommand{\tensor}[1]{\bm{\mathsf{#1}}}

\begin{document}

%\preprint{PREPRINT}
\begin{frontmatter}
\title{Local Vorticity Computation in Double Distribution Functions based Lattice Boltzmann Methods for Flow and Scalar Transport}
\author[label1]{Farzaneh  Hajabdollahi}
\ead{farzaneh.hajabdollahiouderji@ucdenver.edu}
\author[label1]{Kannan N. Premnath}
\ead{kannan.premnath@ucdenver.edu}
\address[label1]{Department of Mechanical Engineering, University of Colorado Denver, 1200 Larimer street, Colorado 80217 , U.S.A}

\date{\today}% It is always \today, today,

\begin{abstract}
Computation of vorticity, or the skew-symmetric velocity gradient tensor, in conjunction with the strain rate tensor, plays an important role in fluid
mechanics in the flow classification, in vortical structure identification and in the modeling of various complex fluids and flows. For the simulation of flows accompanied by the advection-diffusion transport of a scalar field, double distribution functions (DDF) based lattice Boltzmann (LB) methods, involving a pair of LB schemes are commonly used. We present a new local vorticity computation approach by introducing an intensional anisotropy of the scalar flux in the third order, off-diagonal moment equilibria of the LB scheme for the scalar field, and then combining the second order non-equilibrium components of both the LB methods. As such, any pair of lattice sets in the DDF formulation that can independently support the third order off-diagonal moments would enable local determination of the complete flow kinematics, with the LB methods for the fluid motion and the transport of the passive scalar respectively providing the necessary moment relationships to determine the symmetric and skew-symmetric components of the velocity gradient tensor. Since the resulting formulation is completely local and do not rely on finite difference approximations for velocity derivatives, it is by design naturally suitable for parallel computation. As an illustration of our approach, we formulate a DDF-LB scheme for local vorticity computation using a pair of multiple relaxation times (MRT) based collision approaches on two-dimensional, nine velocity (D2Q9) lattices, where the necessary moment relationships to determine the velocity gradient tensor and the vorticity are established via a Chapman-Enskog analysis. Simulations of various benchmark flows demonstrate good accuracy of the predicted vorticity fields using our approach against available solutions, including numerical results, with a second order convergence. Furthermore, extensions of our formulation for a variety of collision models to enable local vorticity computation are presented.
\end{abstract}
\begin{keyword}
Vorticity, Skew-symmetric velocity gradient tensor, Fluid Flow, Scalar Transport, Lattice Boltzmann method, Complex fluids \end{keyword}

\end{frontmatter}

\section{\label{app:Intro}Introduction}
Qualitative distribution and quantitative measures of vorticity is of fundamental interest in fluid mechanics. Indeed, fluid motions
are often associated with vortical structures, which can be characterized by vorticity, and, more generally, by certain invariants of
the velocity gradient tensor~\cite{saffman1992vortex,wu2007vorticity}. The significance of the rigid-body like rotational component of
the fluid element was first identified in a pioneering work by Helmholtz~\cite{helmholtz1867lxiii} and the subject has a long and
rich history~\cite{aref2010150,truesdell2018kinematics}. This local rotational property of the flow, given by the curl of the velocity
field, was termed vorticity by Lamb~\cite{lamb1993hydrodynamics}. While there is no consensus on a rigorous definition of a vortex,
various quantitative measures have been devised to identify regions associated with more rigid-body like rotations than stretching or shearing
motions that aid in flow classification~\cite{truesdell1953two,hunt1988eddies,chong1990general,jeong1995identification,zhou1999mechanisms,chakraborty2005relationships,haller2005objective,zhang2006eigen,kolavr2007vortex,haller2016defining,elsas2017vortex,gao2018rortex,tian2018definitions}.
Such approaches are based on a complete knowledge of the velocity gradient tensor, and the local, Eulerian based methods for coherent structure identification are popular (see~\cite{epps2017review} for recent review).

In more detail, the velocity gradient tensor $A_{ij}\equiv \partial_j u_i$ of the velocity field $u_i$ can be decomposed into symmetric $S_{ij}$ and anti- or skew-symmetric parts $\Omega_{ij}$ as
\begin{equation}
\partial_j u_i =\frac{1}{2}(\partial_j u_i + \partial_i u_j) + \frac{1}{2}(\partial_j u_i - \partial_i u_j) = S_{ij} + \Omega_{ij},
\end{equation}
where $S_{ij}$ is the strain rate tensor and $\Omega_{ij}$ is the intrinsic rotation rate (spin) tensor, with $\Omega_{ij}=-\frac{1}{2}\epsilon_{ijk}\omega_k$. Here, $\omega_k$ is the Cartesian component of the vorticity and $\epsilon_{ijk}$ is the Levi-Civita (permutation) tensor, and the vorticity can be defined as $\omega_i=\epsilon_{ijk}\partial_j u_k$ or $\bm{\omega}=\bm{\nabla}\times \bm{u}$. Both $\omega_i$ and $S_{ij}$, or, in general, $\partial_j u_i$ play an important role in eduction techniques for vortex structure identification. In particular, many of these methods~\cite{epps2017review} are based on the second and third invariants of the velocity gradient tensor $\partial_j u_i$, i.e., $Q=-\frac{1}{2}S_{ij}S_{ij}+\frac{1}{4}\omega_k\omega_k$ and $R=\frac{1}{3}(S_{ij}S_{jk}S_{ki}+\frac{3}{4}\omega_i\omega_jS_{ij})$. Similarly, sometimes
the Lamb vector $L_i = \epsilon_{ijk}\omega_ju_k$ plays a prominent role in the analysis of vortex dynamics~\cite{hamman2008lamb}. Thus, a complete knowledge of the local velocity gradient tensor $\partial_j u_i$, or equivalently, $S_{ij}$ and $\Omega_{ij}$ or $\omega_k$ is of basic interest in
structure identification and classification of flows. This also allows a local determination of the components of the convective acceleration of the fluid elements. In addition, the distribution of vorticity is related to the sound generation and propagation in flow generated acoustics~\cite{howe2003theory}. Furthermore, many models for the representation of turbulence (e.g.,~\cite{pope1975more}), rheological fluid flows such as those involving viscoelasticity, and complex fluid systems such as liquid crystals and polar fluids depend on the local measures of the complete velocity gradient tensor $\partial_j u_i$~\cite{leslie1979theory,beris1994thermodynamics,larson1999structure,deville2012mathematical}. Moreover, molecular liquid flows under nanoscale confinement involves the relaxation of the intrinsic angular momentum to the vorticity of the fluid element, and its coupling to the linear momentum, which needs to be modeled~\cite{de2013non,hansen2009molecular,hansen2011nanoflow,hansen2015continuum}. It is thus highly desirable for computational methods for fluid dynamics that allow especially local determination of all components of the velocity gradient tensor, including the skew symmetric part (i.e., the vorticity). Here, we emphasize that `local' implies that such methods do not depend on finite difference approximations for velocity derivatives, but are entirely based on operations of suitable quantities available at the grid nodes, and hence are naturally suitable for parallel computing.

The lattice Boltzmann method (LBM) is a kinetic computational approach for a variety of fluid mechanics and transport problems
~\cite{He1997,dHumieres2002,Succi2001,Aidun2010,luo2010lattice,Guo2013,Geier2015,Kruger2016}. Generally, the standard versions of the LB schemes can only
represent the symmetric part of $\partial_j u_i$, i.e., the strain rate tensor $S_{ij}$ based on local algorithms via the second order non-equilibrium moments of the distribution function, which are, in turn, related to the spatial derivatives of the first and third order moment equilibria. The latter are constructed based on symmetry and isotropy considerations that respect the underlying isotropy of the viscous stress tensor of the fluid motion represented by the Navier-Stokes equations. It is known that such LB approaches can recover the strain rate tensor components locally with second order accuracy (see e.g., ~\cite{kruger2010second,yong2012accuracy,Ning2016}). However, most of the existing LBMs are not constructed to recover the antisymmetric velocity gradient tensor $\Omega_{ij}$ locally and need to rely on finite difference computations. One notable exception is the recent and interesting work~\cite{peng2017lattice}, which introduced an approach based on modifying the fifth order moment equilibria of the LB solver for fluid flow that enables vorticity computation. This approach is restricted to only lattices that can support fifth order independent moments and thus is applicable only to the three-dimensional, twenty seven velocity (D3Q27) lattice, and not for other standard lattice sets, including the common two-dimensional, nine velocity (D2Q9) lattice, and D3Q15 and D3Q19 lattices in 3D. Furthermore, since it is based on certain prescribed form of the higher order moment equilibria, it may be challenging to extend such LB scheme for thermal flows as well as those with significant compressibility effects that involve constraints on the higher moments of the single distribution function, and may also impact its Galilean invariance of solving the fluid motion. Also, since it involves combining second and fourth order non-equilibrium moments, which may be subjected to hyperviscosity effects~\cite{Geier2015}, the attendant higher order moment equilibria for the solution of the fluid motion need to be constructed carefully.

Our approach is based on different considerations than the above mentioned work for vorticity computation. When the goal is to simulate the fluid motion along with an advection-diffusion transport of a scalar field, represented by the following Navier-Stokes equations (NSE) and the convection-diffusion equation (CDE), respectively:
\begin{eqnarray}
\partial_t \rho + \bm{\nabla}\cdot (\rho \bm{u})&=&0,\label{eq:continuityeqn}\\
\partial_t (\rho \bm{u}) + \bm{\nabla}\cdot (\rho\bm{u}\bm{u})&=&-\bm{\nabla} p + \bm{\nabla}\cdot \tensor{T} + \bm{F},\label{eq:momentumeqn}\\
\partial_t \phi +\bm{\nabla}\cdot (\phi \bm{u})&=&\bm{\nabla}\cdot (D_\phi \bm{\nabla}\phi),\label{eq:scalarfieldeqn}
\end{eqnarray}
where $\rho$, $\bm{u}$ and $p$ are the fluid density, velocity, and pressure, respectively, $T_{ij}=2\rho\nu S_{ij}-\frac{2}{d}\rho\nu\partial_ku_k\delta_{ij}+\rho\zeta\partial_ku_k\delta_{ij}$ is the deviatoric stress tensor (with $\nu$ and $\zeta$ being the kinematic shear and bulk viscosities, respectively, and $d$ being the number of spatial dimensions), $\bm{F}$ is the local body force, and $\phi$ is the scalar field (with $D_\phi$ being its diffusivity), they can be solved by means of a double distribution functions (DDF) based approach using two LB schemes -- one for the flow field and the other for the scalar field. Such situations related to solving the additional passive scalar
field dynamics arise widely, including those related to the transport of energy or temperature field in thermal convection, and of the concentration field of a chemical species in reacting systems, as well as in the interface capturing using phase field models in multiphase flows. Indeed, the modeling of flow and scalar transport using DDF based LBEs is quite common and is a subject of a number of investigations (e.g.,~\cite{ponce1993lattice,he1998novel,van2000convection,lallemand2003theory,rasin2005multi,chopard2009lattice,yoshida2010multiple,wang2013lattice,Chai2013,Contrino2014,Hajabdollahi2018,hajabdollahi2018symmetrized,hajabdollahi2019cascaded}). In such cases, our essential philosophy is to use the additional degrees of freedom (DOF) available in the LBE for the solution of the CDE to construct a procedure for local vorticity computation~\cite{Hajabdollahiphdthesis}. This is possible because as the evolution of the scalar field $\phi$ is influenced by the local fluid velocity $\bm{u}$, its solution procedure can, in principle, contain the complete kinematics of the flow field, which can be obtained from the corresponding LBM with careful construction of its equilibria.

The basic idea behind our approach is as follows. Local vorticity computation in the DDF-LB schemes can be achieved by prescribing an intensional anisotropy of the scalar flux in the third order, off-diagonal moment equilibria of the LBM for the scalar field and then combining the second order, off-diagonal non-equilibrium moment components of both the LBMs. In essence, the LBM for the fluid flow provides local expressions for the strain rate tensor $S_{ij}$ and the LBM for the scalar field yields local relations for the skew-symmetric velocity gradient tensor $\Omega_{ij}$, and hence the vorticity $\omega_k$. This formulation leads to various advantages. The numerical characteristics of the LBM for the fluid motion are preserved as no additional modifications in terms of constraints on its equilibria are imposed (but only on those for the scalar field) and the resulting approach is non-invasive in representing the fluid flow. The freedom from the need to prescribing extra constraints for higher moments for the LB flow solver allows ready extension to construct LB schemes for complex flow physics. In addition, any pair of lattice sets, each supporting only lower (i.e., third) order independent moments, in this DDF-LBE approach can enable local vorticity computation. Thus, this method is applicable for all standard lattices (e.g., D2Q9, D3Q15, D3Q19 and D3Q27) and in different dimensions. Furthermore, the since method is based on two distribution functions which by themselves are generally solved with second order accuracy, the numerically predicted vorticity magnitudes are second order by construction, just like the computed strain rate tensor. Moreover, the local expressions for the vorticity field, which are not dependent on finite difference approximations of the velocity field, naturally lend themselves to parallel computation. Finally, it can be used to establish a LB framework for fully local modeling and computation of complex fluids (e.g., viscoelastic or polar fluids), which generally depend on both the symmetric and skew-symmetric velocity gradient tensor and are usually represented by the evolution of additional distribution functions to represent the attendant multiphysics effects.

For the purpose of illustration without losing generality, in this work, we will specialize our DDF approach by formulating in detail two LB schemes using natural (non-orthogonal) moment basis and multiple relaxation times (MRT) for the solution of flow and scalar transport using the standard D2Q9 lattice to locally compute the complete information about the flow kinematics, including the skew-symmetric velocity gradient tensor components. However, our method can be readily extended to LBM based on other collision models and various other lattice sets in different dimensions. For completeness, we will also present the extension of our approach for the single relaxation time (SRT) LBM and the cascaded central moment LB schemes for the D2Q9 lattice in the appendices. While the objective of this paper is to formulate, mathematically analyze and numerical validate our new LB approach in 2D, its extension to 3D lattices will be presented in a follow-up investigation.

This paper is organized as follows. The next section (Sec.~\ref{sec:LBEfluidmotion}) will present a MRT-LBM for computing the fluid motion, and its Chapman-Enskog (C-E) analysis to determine the symmetric components of the velocity gradient tensor. Section~\ref{sec:LBEscalartransport} will then discuss another MRT-LBM for representing the advection-diffusion transport of a scalar field with the required modifications as indicated earlier, and its C-E analysis to obtain the necessary relations for the skew-symmetric components of the velocity gradient tensor. The expression for the local computation of the vorticity field is derived in Sec.~\ref{sec:vorticitycomputation}. Then, results and discussion of the comparisons of the computed vorticity fields against the analytical/numerical solutions for various representative fluid flow problems are given in Sec.~\ref{sec:resultsanddiscussion}. Finally, Sec.~\ref{sec:summaryandconclusions} presents a summary and conclusions of this work. In addition,~\ref{app:relation_noneqmoments_spatial_derivatives_eqm_moments} presents the system of non-equilibrium moments and spatial derivatives of various attendant components of moment equilibria needed in the derivation of our approach.~\ref{app:SRT_LBM_CDE} discusses a formulation to recover the skew-symmetric velocity gradient tensor for the SRT-LBM, while~\ref{app:Cascaded_LBM_CDE} and~\ref{app:Central_Moment_LBM_CDE} present extensions of our idea for different versions of the LBM based on central moments.

\section{\label{sec:LBEfluidmotion}MRT-LBM for Fluid Motion}
In order to solve the fluid motion in two-dimensions (2D) represented by the mass and momentum conservation equations given in Eqs.~(\ref{eq:continuityeqn}) and (\ref{eq:momentumeqn}), respectively, we will now present a MRT-LBM using a natural, non-orthogonal moment basis~\cite{Premnath2009b}. In this regard, a D2Q9 lattice is used, and whose particle velocities are given by the following:
\begin{subequations}
\begin{eqnarray}
&\ket{e_{ x}} =\left(     0,     1,    0,     -1,     0,  1,-1,-1,1  \right)^\dag,\label{eq:particlevelocities1}\\
&\ket{e_{ y}} =\left(     0,     0,     1,     0,    -1,1,1,-1,-1
\right)^\dag,\label{eq:particlevelocities2}
\end{eqnarray}
\end{subequations}
where $\dag$ is the transpose operator and the standard Dirac's bra-ket notation is used to represent the vectors. The Cartesian components for any particle direction $\alpha$ are represented by $e_{\alpha x}$ and $e_{\alpha y}$, where $\alpha=0,1,\ldots, 8$. In addition, we need the following 9-dimensional vector whose inner product with the particle distribution function $f_\alpha$ yields its zeroth moment:
\begin{eqnarray}
&\ket{1} =\left(     1,     1,    1,     1,     1,  1,1,1,1  \right)^\dag.\label{eq:unit9dimensionalvector}
\end{eqnarray}
The non-orthogonal basis vectors can then be written as
\begin{eqnarray}
&&{{T}_{0}}=\ket{1}, \quad{{T}_{1}}=\ket{e_{ x}},\quad{{T}_{2}}=\ket{e_{ y}}, \quad {{T}_{3}}=\ket{e_{ x}^2+e_{ y}^2}, \quad {{T}_{4}}=\ket{e_{ x}^2-e_{ y}^2}, \nonumber \\
&&{{T}_{5}}=\ket{e_{ x}e_{ y}},\quad {{T}_{6}}=\ket{e_{ x}^2e_{ y}}, \quad {{T}_{7}}=\ket{e_{ x}e_{ y}^2},\quad {{T}_{8}}=\ket{e_{ x}^2e_{ y}^2}.\label{eq:basisvectors}
\end{eqnarray}
In the above, symbols such as $\ket{e_{ x}^2e_{ y}}=\ket{e_{ x}e_{ x}e_{ y}}$ denote a vector that arise from the elementwise vector multiplication of vectors $\ket{e_{ x}}$, $\ket{e_{ x}}$ and $\ket{e_{ y}}$. In order to map changes of moments back to changes in the distribution function, we group the above set of vectors as a transformation matrix $\tensor{T}$, which reads as
\begin{equation}
\tensor{T}=\left[{{T}_{0}},{{T}_{1}},{{T}_{2}},{{T}_{3}},{{T}_{4}},{{T}_{5}},{{T}_{6}},{{T}_{7}},{{T}_{8}}\right].
\label{eq:transformationmatrix}
\end{equation}

We then define the raw moments of order ($m+n$) of the distribution function $f_\alpha$, its equilibrium $f_\alpha^{eq}$, and the source terms $S_{\alpha}$ to represent the body force, respectively, as
\begin{eqnarray}
\left( {\begin{array}{*{20}{l}}
{{{\hat \kappa }_{{x^m}{y^n}}}}^{'}\\
{\hat {\kappa} _{{x^m}{y^n}}^{eq'}}\\
{\hat {\sigma} _{{x^m}{y^n}}^{eq'}}
\end{array}} \right) = \sum\limits_{\alpha=0}^8  {\left( {\begin{array}{*{20}{l}}
{{f_\alpha }}\\
{f_\alpha ^{eq}}\\
{S_\alpha}
\end{array}} \right)} {e_{\alpha x}^m}{e_{\alpha y}^n}.
\label{eq:rawmomentdefinitions}
\end{eqnarray}
Here, and in what follows, the prime ($'$) symbols denote various raw moments. In terms of the nominal, nonorthogonal transformation matrix $\tensor{T}$ the relation between the various moments and their corresponding states in the velocity space can be written as
\begin{equation}
\mathbf{\widehat{m}}=\tensor{T}\mathbf{f},\quad
\mathbf{\widehat{m}}^{eq}=\tensor{T}\mathbf{f}^{eq}, \quad
\mathbf{\widehat{S}}=\tensor{T}\mathbf{S},
\label{eq:relationvelocityspace_to_momentspace}
\end{equation}
where
\begin{eqnarray*}
\mathbf{{f}}&=&\left({f}_{0},{f}_{1},{f}_{2},\ldots,{f}_{8}\right)^{\dag}, \quad \mathbf{{f}}^{eq}=\left({f}_{0}^{eq},{f}_{1}^{eq},{f}_{2}^{eq},\ldots,{f}_{8}^{eq}\right)^{\dag}, \\ \mathbf{{S}}&=&\left({S}_{0},{S}_{1},{S}_{2},\ldots,{S}_{8}\right)^{\dag}
\end{eqnarray*}
are the various quantities in the velocity space, and
\begin{subequations}
\begin{eqnarray}
\mathbf{\widehat{m}}&=&\left(\widehat{m}_{0},\widehat{m}_{1},\widehat{m}_{2},\ldots,\widehat{m}_{8}\right)^{\dag}\nonumber\\
&=&\left(\widehat{\kappa}_{0}^{'},\widehat{\kappa}_{x}^{'},\widehat{\kappa}_{y}^{'},
\widehat{\kappa}_{xx}^{'}+\widehat{\kappa}_{yy}^{'},\widehat{\kappa}_{xx}^{'}-\widehat{\kappa}_{yy}^{'},
\widehat{\kappa}_{xy}^{'},\widehat{\kappa}_{xxy}^{'},\widehat{\kappa}_{xyy}^{'},
\widehat{\kappa}_{xxyy}^{'}\right)^{\dag},\\
\mathbf{\widehat{m}}^{eq}&=&\left(\widehat{m}_{0}^{eq},\widehat{m}_{1}^{eq},\widehat{m}_{2}^{eq},\ldots,\widehat{m}_{8}^{eq}\right)^{\dag}\nonumber\\
&=&\left(\widehat{\kappa}_{0}^{eq'},\widehat{\kappa}_{x}^{eq'},\widehat{\kappa}_{y}^{eq'},
\widehat{\kappa}_{xx}^{eq'}+\widehat{\kappa}_{yy}^{eq'},\widehat{\kappa}_{xx}^{eq'}-\widehat{\kappa}_{yy}^{eq'},
\widehat{\kappa}_{xy}^{eq'},\widehat{\kappa}_{xxy}^{eq'},\widehat{\kappa}_{xyy}^{eq'},
\widehat{\kappa}_{xxyy}^{eq'}\right)^{\dag},\\
\mathbf{\widehat{S}}&=&\left(\widehat{S}_{0},\widehat{S}_{1},\widehat{S}_{2},\ldots,\widehat{S}_{8}\right)^{\dag}\nonumber\\
&=&\left(\widehat{\sigma}_{0}^{'},\widehat{\sigma}_{x}^{'},\widehat{\sigma}_{y}^{'},
\widehat{\sigma}_{xx}^{'}+\widehat{\sigma}_{yy}^{'},\widehat{\sigma}_{xx}^{'}-\widehat{\sigma}_{yy}^{'},
\widehat{\sigma}_{xy}^{'},\widehat{\sigma}_{xxy}^{'},\widehat{\sigma}_{xyy}^{'}
\widehat{\sigma}_{xxyy}^{'}\right)^{\dag}
\end{eqnarray}
\end{subequations}
are the corresponding states in the moment space.

The MRT-LBM with trapezoidal rule to represent the source term with second order accuracy can be written as
\begin{eqnarray}
\mathbf f\left( {\bm{x} + {{ \bm e}_\alpha }{\delta _t},t + {\delta _t}} \right) - \mathbf f\left( {\bm x,t} \right) &=& {\tensor{T}^{ - 1}}\left[ { - \hat {\tensor\Lambda} \left( {\widehat {\mathbf m} - {{\widehat {\mathbf m}}^{eq}}} \right)} \right] \nonumber\\
&+& \frac{1}{2}{\tensor{T}^{ - 1}}\left[ \widehat {\mathbf{S}}\left( {\bm{x} + {{ \bm e}_\alpha }{\delta _t},t + {\delta _t}} \right) + \widehat {\mathbf{S}}\left( {\bm{x},t}\right)\right]{\delta _t},
\label{eq:MRT-LBEfluidmotion}
\end{eqnarray}
where the diagonal relaxation time matrix $\hat {\tensor \Lambda}$ can be represented as
\begin{equation}
\hat {\tensor \Lambda}=\mbox{diag}(0,0,0,\omega_3,\omega_4,\omega_5,\omega_6,\omega_7,\omega_8).
\label{eq:relaxationtimematrix_fluidmotion}
\end{equation}
In order to obtain an effectively explicit scheme, we apply the transformation~\cite{He1998b,He1999} $\bar{f}_{\alpha}=f_{\alpha}-\frac{1}{2}S_{\alpha}\delta_t$, or equivalently $\widehat{\bar{\mathbf{m}}}=\widehat{\mathbf{m}}-\frac{1}{2}\widehat{\mathbf{S}}\delta_t$ and $\widehat{\bar{\kappa}}_{{x^m}{y^n}}^{'}={\hat \kappa }_{{x^m}{y^n}}^{'}-\frac{1}{2}{\hat \sigma }_{{x^m}{y^n}}^{'}\delta_t$, and the MRT-LBE can be written as
\begin{equation}
\begin{aligned}
\bar{\mathbf{f}}\left( {\bm{x} + {{ \bm e}_\alpha }{\delta _t},t + {\delta _t}} \right) - \bar{\mathbf{f}}\left( {\bm x,t} \right) = {\tensor{T}^{ - 1}}\left[ { - \hat {\tensor\Lambda} \left( {\widehat {\bar{\mathbf{m}}} - {{\widehat {\mathbf m}}^{eq}}} \right)} \right] + {\tensor{T}^{ - 1}}\left[ {\left( {\tensor{I} - \frac{1}{2}\hat { \tensor\Lambda} } \right)\widehat {\mathbf S}} \right]{\delta _t},
\label{eq:MRT-LBEfluidmotion_transformed}
\end{aligned}
\end{equation}
The moment equilibria $\hat {\kappa} _{{x^m}{y^n}}^{eq'}$ at different orders can be written as~\cite{Premnath2009b}
\begin{eqnarray}
&\widehat{\kappa}^{eq'}_{0}=\rho,\
\widehat{\kappa}^{eq'}_{x}=\rho u_x, \
\widehat{\kappa}^{eq'}_{y}=\rho u_y, \nonumber\\
&\widehat{\kappa}^{eq'}_{xx}=c_s^2\rho+\rho u_x^2,\
\widehat{\kappa}^{eq'}_{yy}=c_s^2\rho+\rho u_y^2,\
\widehat{\kappa}^{eq'}_{xy}=\rho u_x u_y, \nonumber\\
&\widehat{\kappa}^{eq'}_{xxy}=c_s^2\rho u_y+\rho u_x^2u_y, \
\widehat{\kappa}^{eq'}_{xyy}=c_s^2\rho u_x+\rho u_xu_y^2,\nonumber\\&
\widehat{\kappa}^{eq'}_{xxyy}=c_s^4\rho+c_s^2\rho (u_x^2+u_y^2)+\rho u_x^2u_y^2,\label{eq:eqmrawmoment}
\end{eqnarray}
which are obtained from the discrete representation of the local Maxwellian by transforming back their central moments at a given order to their corresponding raw moments. Here, $c_s$ is the speed of sound, and in the present work, we typically set $c_s^2=1/3$. Also, moments of the source terms $\hat {\sigma} _{{x^m}{y^n}}^{eq'}$ follows as~\cite{Premnath2009b}
\begin{eqnarray}
&\widehat{\sigma}^{'}_{0}=0,\
\widehat{\sigma}^{'}_{x}=F_x,\
\widehat{\sigma}^{'}_{y}=F_y, \nonumber\\
&\widehat{\sigma}^{'}_{xx}=2F_xu_x,\
\widehat{\sigma}^{'}_{yy}=2F_yu_y,\
\widehat{\sigma}^{'}_{xy}=F_xu_y+F_yu_x,\nonumber\\
&\widehat{\sigma}^{'}_{xxy}=F_yu_x^2+2F_xu_xu_y,\
\widehat{\sigma}^{'}_{xyy}=F_xu_y^2+2F_yu_yu_x,\nonumber\\&
\widehat{\sigma}^{'}_{xxyy}=2(F_xu_xu_y^2+F_yu_yu_x^2),
\end{eqnarray}
where $\bm{F}=(F_x,F_y)$. The hydrodynamic fields are given by
\begin{equation}\label{eq:hydrodynamic_fields}
\rho = \sum_{\alpha=0}^8 \bar{f}_{\alpha}, \quad \rho \bm{u} = \sum_{\alpha=0}^8 \bar{f}_{\alpha}\bm{e}_{\alpha}+\frac{1}{2}\bm{F}\delta_t, \quad p = c_s^2\rho,
\end{equation}
where $\bm{u}=(u_x,u_y)$. The above represents the solution of the NSE (Eqs.~(\ref{eq:continuityeqn}) and (\ref{eq:momentumeqn})), with the kinematic bulk and shear viscosities related to the relaxation times via $\zeta=c_s^2\left(\frac{1}{\omega_3}-\frac{1}{2}\right)\delta_t$ and $\nu=c_s^2\left(\frac{1}{\omega_j}-\frac{1}{2}\right)\delta_t$, where $j=4,5$ respectively. The remaining relaxation times for the higher order moments, which influence the numerical stability, are set to unity in this work.

\subsection{\label{subsec:CEanalysis_symmvelgradtensor} Moment relationships for the symmetric velocity gradient tensor: Chapman-Enskog Analysis}
We will now perform a Chapman-Enskog analysis~\cite{chapman1990mathematical} to determine the expressions that relate the symmetric velocity gradient tensor to certain components of the local (non-equilibrium) moments. Expanding the moments about its equilibria as well as applying the standard multiscale expansion of the time derivatives in the MRT-LB scheme given in the previous section
\begin{equation}
\mathbf{\widehat{m}}=\sum_{j=0}^{\infty}\epsilon^j \mathbf{\widehat{m}}^{(j)}, \quad \partial_t=\sum_{j=0}^{\infty}\epsilon^j \partial_{t_j},\label{eq:CE_expansions}
\end{equation}
where $\epsilon$ is ${a}$ small bookkeeping perturbation parameter, and also performing a Taylor series expansion of the streaming operator in Eq.~(\ref{eq:MRT-LBEfluidmotion_transformed}), i.e.,
\begin{equation}
\bar{\mathbf{f}}(\bm{x}+\bm{e}_{\alpha}\epsilon,t+\epsilon)=\sum_{j=0}^{\infty}\frac{\epsilon^j}{j!}(\partial_t+\bm{e}_{\alpha}\cdot\bm{\nabla})^j\bar{\mathbf{f}}(\bm{x},t).
\label{eq:11}
\end{equation}
and converting all quantities in the velocity space to the moment space (via Eq.~(\ref{eq:relationvelocityspace_to_momentspace})) and using $\widehat{\bar{\mathbf{m}}}=\widehat{\mathbf{m}}-\frac{1}{2}\widehat{\mathbf{S}}\delta_t$, we obtain the following system of moment equations at consecutive order in $\epsilon$:
\begin{subequations}
\begin{eqnarray}
&O(\epsilon^0):\quad \mathbf{\widehat{m}}^{(0)}=\mathbf{\widehat{m}}^{eq},\label{eq:CEmoment0thfluidmotion}\\
&O(\epsilon^1):\quad (\partial_{t_0}+\widehat{\tensor E}_i \partial_i)\mathbf{\widehat{m}}^{(0)}=-\widehat{\tensor\Lambda}\mathbf{\widehat{m}}^{(1)}+\mathbf{\widehat{S}},\label{eq:CEmoment1stfluidmotion}\\
&O(\epsilon^2):\quad \partial_{t_1}\mathbf{\widehat{m}}^{(0)}+(\partial_{t_0}+\widehat{\tensor E}_i \partial_i)\left[ \tensor{I}-\frac{1}{2}\widehat{\tensor\Lambda}\right]\mathbf{\widehat{m}}^{(1)}=-\widehat{\tensor\Lambda}\mathbf{\widehat{m}}^{(2)},\label{eq:CEmoment2ndfluidmotion}
\end{eqnarray}
\end{subequations}
where $\widehat{\tensor E}_i=\tensor{T}(\bm e_{ i}\tensor{I})\tensor{T}^{-1}, i \in \{x,y\}$. In order to obtain the hydrodynamic macroscopic equations, in the leading, i.e., $O(\epsilon)$ system (see Eq.~(\ref{eq:CEmoment1stfluidmotion})), the equations representing the evolution of the moment components up to the second order are necessary, which read as (see~\ref{app:relation_noneqmoments_spatial_derivatives_eqm_moments} for details)
\begin{subequations}
 \begin{eqnarray}
&\partial_{t_0}\rho+\partial_x (\rho u_x)+\partial_y (\rho u_y) = 0,\label{eq:1stmoment_fluidmotion_a}\\&
\partial_{t_0}\left(\rho u_x\right)+\partial_x \left({c_s^2\rho}+\rho u_x^2\right)+\partial_y \left(\rho u_xu_y\right) = F_x,
\label{eq:1stmoment_fluidmotion_b}\\&
\partial_{t_0}\left(\rho u_y\right)+\partial_x \left(\rho u_xu_y\right)+\partial_y \left(c_s^2\rho+\rho u_y^2\right) = F_y,
\label{eq:1stmoment_fluidmotion_c}\\&
\partial_{t_0}\left(2c_s^2\rho+\rho(u_x^2+u_y^2)\right)+\partial_x \left[(1+c_s^2)\rho u_x+\rho u_xu_y^2\right]+\partial_y \left[(1+c_s^2)\rho u_y+\rho u_x^2u_y\right]\nonumber\\& =-\omega_3\widehat{m}_3^{(1)}+2({F_xu_x}+{F_yu_y}),
\label{eq:1stmoment_fluidmotion_d}\\&
\partial_{t_0}\left(\rho(u_x^2-u_y^2)\right)+\partial_x \left[(1-c_s^2)\rho u_x-\rho u_xu_y^2\right]+\partial_y \left[(-1+c_s^2)\rho u_y+\rho u_x^2u_y\right]\nonumber\\&
=-\omega_4\widehat{m}_4^{(1)}+2(F_xu_x-F_yu_y),
\label{eq:1stmoment_fluidmotion_e}\\&
\partial_{t_0}\left(\rho u_x u_y\right)+\partial_x \left(c_s^2\rho u_y+\rho u_x^2u_y\right)+\partial_y \left(c_s^2\rho u_x+\rho u_xu_y^2\right)\nonumber\\&
 =-\omega_5\widehat{m}_5^{(1)}+F_xu_y+F_yu_x.
\label{eq:1stmoment_fluidmotion_f}
\end{eqnarray}
\end{subequations}
Analogously, at the next, i.e., $O(\epsilon^2)$ level (see Eq.~(\ref{eq:CEmoment2ndfluidmotion})), the relevant moment equations to recover the equations of the fluid motion written up to the first order as
\begin{subequations}
\begin{eqnarray}
&\partial_{t_1}\rho=0,
\label{eq:2ndmoment_fluidmotion_a}\\&
\partial_{t_1}\left(\rho u_x\right)+\partial_x \left[\frac{1}{2}\left(1-\frac{1}{2}\omega_3\right)\widehat{m}_3^{(1)}+\frac{1}{2}\left(1-\frac{1}{2}\omega_4\right)\widehat{m}_4^{(1)}\right]
+\partial_y \left[\left(1-\frac{1}{2}\omega_5\right)\widehat{m}_5^{(1)}\right]\nonumber \\&=0,
\label{eq:2ndmoment_fluidmotion_b}\\&
\partial_{t_1}\left(\rho u_y\right)+\partial_x \left[\left(1-\frac{1}{2}\omega_5\right)\widehat{m}_5^{(1)}\right]+\partial_y \left[\frac{1}{2}\left(1-\frac{1}{2}\omega_3\right)\widehat{m}_3^{(1)}-\frac{1}{2}\left(1-\frac{1}{2}\omega_4\right)\widehat{m}_4^{(1)}\right]\nonumber \\
&=0.
\label{eq:2ndmoment_fluidmotion_c}
\end{eqnarray}
\end{subequations}
Here, the components of the second-order non-equilibrium moments $\widehat{m}_3^{(1)}$, $\widehat{m}_4^{(1)}$ and $\widehat{m}_5^{(1)}$ (which represent $\widehat{\kappa}_{xx}^{\prime(1)}+\widehat{\kappa}_{yy}^{\prime(1)}$, $\widehat{\kappa}_{xx}^{\prime(1)}-\widehat{\kappa}_{yy}^{\prime(1)}$ and $\widehat{\kappa}_{xy}^{\prime(1)}$, respectively) are unknowns. They can be obtained from Eqs.~(\ref{eq:1stmoment_fluidmotion_d}), (\ref{eq:1stmoment_fluidmotion_e}) and (\ref{eq:1stmoment_fluidmotion_f}), respectively, where the time derivatives $\partial_{t_0}\left(2c_s^2\rho+\rho(u_x^2+u_y^2)\right)$, $\partial_{t_0}\left(\rho(u_x^2-u_y^2)\right)$ and $\partial_{t_0}\left(\rho u_x u_y\right)$          are eliminated in favor the spatial derivatives using the leading order mass and momentum equations (i.e., Eqs.~(\ref{eq:1stmoment_fluidmotion_a})--(\ref{eq:1stmoment_fluidmotion_c}), respectively). For details, see e.g.,~\cite{Premnath2009b,hajabdollahi2018galilean}. Neglecting all terms of $O(u^3)$ and higher, we can obtain the expressions for the various components of the non-equilibrium second order moments related to the symmetric part of the velocity gradient tensor $S_{ij}=\frac{1}{2}(\partial_ju_i+\partial_iu_j)$ (i.e., $\partial_xu_x$, $\partial_yu_y$ and $\partial_yu_x+\partial_yu_x$), which read as~\cite{Premnath2009b,hajabdollahi2018galilean}
\begin{subequations}
\begin{eqnarray}
\widehat{m}_3^{(1)}=\widehat{\kappa}_{xx}^{(1)'}+\widehat{\kappa}_{yy}^{(1)'}&=&-\frac{2c_s^2\rho}{\omega_3 }(\partial_xu_x+\partial_yu_y), \label{eq:strainratexxpyy}\\
\widehat{m}_4^{(1)}=\widehat{\kappa}_{xx}^{(1)'}-\widehat{\kappa}_{yy}^{(1)'}&=&-\frac{2c_s^2\rho}{\omega_4 }(\partial_xu_x-\partial_yu_y),\label{eq:strainratexxmyy}\\
\widehat{m}_5^{(1)}=\widehat{\kappa}_{xy}^{(1)'}&=&-\frac{c_s^2\rho}{\omega_5 }(\partial_xu_y+\partial_yu_x). \label{eq:strainratexy}
\end{eqnarray}
\end{subequations}
When these expressions are substituted in Eqs.~(\ref{eq:2ndmoment_fluidmotion_b}) and (\ref{eq:2ndmoment_fluidmotion_c}), and then combining the $O(\epsilon)$ and $O(\epsilon^2)$ moment equations up to the first order, the NSE given Eqs.~(\ref{eq:continuityeqn}) and (\ref{eq:momentumeqn}) follows. The non-equilibrium moment relations given in Eqs.~(\ref{eq:strainratexxpyy})--(\ref{eq:strainratexy}) will be combined further with the developments given in the next section to develop a local computing approach for the vorticity field later in Sec.~\ref{sec:vorticitycomputation}.

\section{\label{sec:LBEscalartransport}MRT-LBM for Transport of a Passive Scalar}
The solution of the advection-diffusion of the passive scalar field $\phi$ given by the CDE in Eq.~(\ref{eq:scalarfieldeqn}) will now be represented by using another MRT-LBM. Considering the D2Q9 lattice again, which, as required, supports the off-diagonal third order moment equilibria independently as noted in the Introduction, we use the same natural moment basis given in Eq.~(\ref{eq:basisvectors}) as well as the resulting transformation matrix $\tensor{T}$ (see Eq.~(\ref{eq:transformationmatrix})). First, we define the relation between the various raw moments and the corresponding distribution function $g_{\alpha}$ and their equilibria $g_{\alpha}^{eq}$ for this MRT-LBE as
\begin{eqnarray}
&\hat {\mathbf  n}=\tensor T \mathbf g, \quad \hat {\mathbf {n}}^{eq}=\tensor T \mathbf {g}^{eq},
\label{eq:relationvelocityspace_to_momentspace_CDE}
\end{eqnarray}
where
\begin{eqnarray}
\mathbf g=(g_0,g_1,g_2\dots g_8)^\dagger, \quad {\mathbf {g}}^{eq}=(g_0^{eq},g_1^{eq},g_2^{eq}\dots g_8^{eq})^\dagger
\label{eq:19}
\end{eqnarray}
are given in the velocity space, and
\begin{eqnarray}
\hat {\mathbf n}&=&(\hat {n}_0,\hat {n}_1,\hat {n}_2\dots \hat {n}_8)^\dagger\nonumber\\
&=&(\hat{\eta}_0^\prime,\hat{\eta}_x^\prime,\hat{\eta}_y^\prime,\hat{\eta}_{xx}^\prime+\hat{\eta}_{yy}^\prime,\hat{\eta}_{xx}^\prime-\hat{\eta}_{yy}^\prime,\hat{\eta}_{xy}^\prime,\hat{\eta}_{xxy}^\prime,\hat{\eta}_{xyy}^\prime,\hat{\eta}_{xxyy}^\prime)^\dagger\label{eq:momentCDE}\\
\hat {\mathbf {n}}^{eq}&=&(\hat {n}_0^{eq},\hat {n}_1^{eq},\hat {n}_2^{eq}\dots \hat {n}_8^{eq})^\dagger\nonumber\\
&=&(\hat{\eta}_0^{eq^\prime},\hat{\eta}_x^{eq^\prime},\hat{\eta}_y^{eq^\prime},\hat{\eta}_{xx}^{eq^\prime}+\hat{\eta}_{yy}^{eq^\prime},\hat{\eta}_{xx}^{eq^\prime}-\hat{\eta}_{yy}^{eq^\prime},\hat{\eta}_{xy}^{eq^\prime},\hat{\eta}_{xxy}^{eq^\prime},\hat{\eta}_{xyy}^{eq^\prime},\hat{\eta}_{xxyy}^{eq^\prime})^\dagger\label{eq:momentequilibriaCDE}
\end{eqnarray}
represent the equivalent states in the moment space. Here, the various sets of raw moments are defined as follows:
\begin{eqnarray}
\left( \begin{array}{l}
{{\hat \eta^{\prime} }_{{x^m}{y^n}}}\\
{\hat \eta}^{eq^\prime} _{{x^m}{y^n}}
\end{array} \right) = \sum\limits_{\alpha=0}^8  {\left( \begin{array}{l}
{g_\alpha }\\
g_\alpha ^{eq}\\
\end{array} \right)}{e_{\alpha x}^m}{e_{\alpha y}^n},\label{eq:rawmomentdefinitions_CDE}
\end{eqnarray}

Then the MRT-LBE using a non-orthogonal moment basis for the solution of the CDE can be written as
\begin{equation}
\mathbf{g}(\bm{x}+\bm{e}_\alpha\delta_t,t+\delta_t)-\mathbf{g}(\bm{x},t)= -\tensor{T}^{-1}[\hat {\mathbf \Lambda}^\phi(\hat {\mathbf {n}}-\hat {\mathbf {n}}^{eq})],\label{eq:MRT-LBECDE}
\end{equation}
where $\hat {\mathbf \Lambda}^\phi$ is the diagonal relaxation time matrix given by
\begin{eqnarray}
\hat {\mathbf \Lambda}^\phi=\mbox{diag}(0,\omega_1^\phi,\omega_2^\phi,\omega_3^\phi,\omega_4^\phi,\omega_5^\phi,\omega_6^\phi,\omega_7^\phi,\omega_8^\phi,)
\label{eq:relaxationtimematrix_CDE}
\end{eqnarray}

A key element in this work is the prescription of the moment equilibria $\hat {\mathbf {n}}^{eq}$ (Eq.~(\ref{eq:momentequilibriaCDE})) used in Eq.~(\ref{eq:MRT-LBECDE}) to enable a local computation of the antisymmetric velocity gradient tensor or the  vorticity field. The passive scalar $\phi$ is advected by the local velocity field $\bm{u}$, and $\phi$ hence its solution procedure, in principle, has a complete information on the kinematics of the fluid elements undergoing a variety of motions when it is carefully designed. As such, most of the components of the moment equilibria $\hat {\mathbf n}^{eq}$ can be constructed in analogy with $\hat {\mathbf m}^{eq}$ given in Eq.~(\ref{eq:eqmrawmoment}), where the density $\rho$ is replaced by the scalar field $\phi$. On the other hand, in view of the above consideration, in order to extract the local intrinsic rotation rate of the fluid element related to the antisymmetric velocity gradient tensor, we prescribe anisotropy in the scalar flux $(\phi \bm u)$ components used in the third order moment equilibria, which, as we shall see in the following, does not affect the recovery of the macroscopic CDE. Thus, we set
\begin{eqnarray}
&\widehat{\eta}^{eq'}_{0}=\phi,\
\widehat{\eta}^{eq'}_{x}=\phi u_x, \
\widehat{\eta}^{eq'}_{y}=\phi u_y, \nonumber\\
&\widehat{\eta}^{eq'}_{xx}=c_{s\phi}^{2}\phi+\phi u_x^2,\
\widehat{\eta}^{eq'}_{yy}=c_{s\phi}^{2}\phi+\phi u_y^2,\
\widehat{\eta}^{eq'}_{xy}={\phi u_x u_y}, \nonumber\\
&\widehat{\eta}^{eq'}_{xxy}=\boxed{\beta_1c_{s\phi}^{2}\phi u_y}+\phi u_x^2 u_y, \
\widehat{\eta}^{eq'}_{xyy}=\boxed{\beta_2c_{s\phi}^{2}\phi u_x}+\phi u_x u_y^2,\nonumber\\&
\widehat{\eta}^{eq'}_{xxyy}=c_{s\phi}^{4}\phi+c_{s\phi}^{2}\phi(u_x^2+u_y^2)+\phi u_x^2u_y^2,\label{eq:eqmrawmoment_CDE}
\end{eqnarray}
where $c_{s\phi}$ is an independent parameter related to the diffusivity $D_{\phi}$ (see below), and we typically set $c_{s\phi}^2=1/3$ in this work. Here, $\beta_1$ and $\beta_2$ are free parameters that prescribe anisotropy on the scalar flux appearing in the third order moment equilibria. Typically, $\beta_1\approx	1$ and $\beta_2\approx 1$, but $\beta_1-\beta_2\neq 0$ , i.e., a small intentional anisotropy is introduced to locally recover the magnitude of the intrinsic rotation rate of the fluid motion (see the following section). The scalar field $\phi$ is then obtained as the zeroth moment of the distribution function $g_{\alpha}$, which evolves according to Eq.~(\ref{eq:MRT-LBECDE}) in the form of the standard collide-and-steam steps:
\begin{equation}
\phi = \sum_{\alpha=0}^8 g_{\alpha}.
\end{equation}
Then, the above represents the solution of the CDE (Eq.~(\ref{eq:scalarfieldeqn})), with the diffusivity related to the relaxation times via $D_{\phi}=c_{s\phi}^2\left(\frac{1}{\omega_j^\phi}-\frac{1}{2}\right)\delta _t$ where $j=1,2$. It may be noted that~\ref{app:SRT_LBM_CDE}--\ref{app:Central_Moment_LBM_CDE} present extensions of our approach to other collision models, including the SRT-LBM and central moment LB formulations.

\subsection{\label{subsec:CEanalysis_scalargrad_antisymmvelgradtensor} Moment relationships for the scalar gradient vector and skew-symmetric velocity gradient tensor: Chapman-Enskog Analysis}
We will now perform a C-E analysis of the MRT-LBE for the passive scalar field. Applying the moment expansion about its equilibria and a multiscale expansion of the time derivative to Eq.~(\ref{eq:MRT-LBECDE})
\begin{eqnarray}
\hat {\mathbf n}=\sum\limits_{j=0}^\infty \epsilon^j \hat {\mathbf n}^{(j)}, \quad \partial_t=\sum\limits_{j=0}^\infty  \epsilon^j \partial_{t j},
\label{eq:24}
\end{eqnarray}
where $\epsilon=\delta_t$ and also using a Taylor expansion of the streaming operator ${\mathbf{g}}_{\alpha}(\bm{x}+\bm{e}_\alpha \epsilon,t+\epsilon)=\sum_{j=0}^\infty\frac{\epsilon^j}{j!}( \partial_t+\bm {e}_{\alpha}\cdot \bm {\nabla})^j{\mathbf{g}}(\bm x,t)$, the following moment equations at consecutive order in $\epsilon$ can be obtained:
\begin{subequations}
\begin{eqnarray}
O(\epsilon^0):\quad \mathbf{\widehat{n}}^{(0)}=\mathbf{\widehat{n}}^{eq},\label{eq:CEmoment0thCDE}\\
O(\epsilon^1):\quad (\partial_{t_0}+\widehat{\tensor{E}}_i \partial_i)\mathbf{\widehat{n}}^{(0)}&=&-\widehat{\tensor{\Lambda}}^\phi\mathbf{\widehat{n}}^{(1)},\label{eq:CEmoment1stCDE}\\
O(\epsilon^2):\quad \partial_{t_1}\mathbf{\widehat{n}}^{(0)}+(\partial_{t_0}+\widehat{\tensor{E}}_i \partial_i)\left[ \tensor{I}-\frac{1}{2}\widehat{\tensor{\Lambda}}^\phi\right]\mathbf{\widehat{n}}^{(1)}&=&-\widehat{\tensor{\Lambda}}^\phi\mathbf{\widehat{n}}^{(2)},\label{eq:CEmoment2ndCDE}
\end{eqnarray}
\end{subequations}
where $\widehat{\tensor{E}}_i$ is the same as that given earlier. Some of the relevant components at the leading order (i.e., $(O(\epsilon))$) of the moment system (see Eq.~(\ref{eq:CEmoment1stCDE})) are given as
\begin{subequations}
\begin{eqnarray}
&\partial_{t_0}\phi+\partial_{x}(\phi u_x)+\partial_{y}(\phi u_y)=0,\label{eq:1stmoment_CDE_a}\\
&\partial_{t_0}(\phi u_x)+\partial_x(c_{s\phi}^2+\phi u_x^2)+\partial_y(\phi u_xu_y)=-\omega_1^\phi \hat {n}^{(1)}_1,\label{eq:1stmoment_CDE_b}\\
&\partial_{t_0}(\phi u_y)+\partial_x(\phi u_xu_y)+\partial_y(c_{s\phi}^2+\phi u_y^2)=-\omega_2^\phi \hat {n}^{(1)}_2,\label{eq:1stmoment_CDE_c}\\
&\partial_{t_0}(2c_{s\phi}^2+\phi(u_x^2+u_y^2))+\partial_x\left[(1+\beta_2 c_{s\phi}^2)\phi u_x+\phi u_x u_y^2\right]+\partial_y\left[(1+\beta_1 c_{s\phi}^2)\phi u_y+\phi u_x^2 u_y\right]\nonumber\\&=-\omega_3^\phi \hat {n}^{(1)}_3,\label{eq:1stmoment_CDE_d}\\
&\partial_{t_0}(\phi(u_x^2-u_y^2))+\partial_x\left[(1-\beta_2c_{s\phi}^2)\phi u_x+\phi u_xu_y^2\right]+\partial_y\left[(-1+\beta_1c_{s\phi}^2)\phi u_y+\phi u_x^2u_y\right]\nonumber\\&=-\omega_4\phi \hat {n}^{(1)}_4,\label{eq:1stmoment_CDE_e}\\
&\partial_{t_0}(\phi u_xu_y)+\partial_x\left[\beta_1c_{s\phi}^2\phi u_y+\phi u_x^2u_y\right]+\partial_y\left[\beta_2c_{s\phi}^2\phi u_x+\phi u_xu_y^2\right]\nonumber\\&=-\omega_5\phi \hat {n}^{(1)}_5,\label{eq:1stmoment_CDE_f}
\end{eqnarray}
\end{subequations}
where the above can be obtained by replacing ${\widehat\kappa_{x^my^n}^{eq'}}$ in the corresponding C-E analysis for the fluid motion with $\hat{\eta}_{x^m y^n}^{eq'}$ (see the previous section and~\ref{app:relation_noneqmoments_spatial_derivatives_eqm_moments} for details) and allowing for the relaxation of the first order moments, since only the scalar field $\phi$ is conserved in the present case. Similarly, the leading component (i.e., the zeroth order) of the moment system at the $O(\epsilon^2)$ level to recover the CDE is obtained from Eqs.~(\ref{eq:CEmoment2ndCDE}) can be written as
\begin{eqnarray}
\partial_{t_1}\phi+\partial_x\left[\left(1-\frac{\omega_1^\phi}{2}\right)\hat {n}^{(1)}_1\right]+\partial_y\left[\left(1-\frac{\omega_2^\phi}{2}\right)\hat{ n}^{(1)}_2\right]=0.\label{eq:2ndmoment_CDE_a}
\end{eqnarray}

Now, in order to derive the CDE, we need to combine Eq.~(\ref{eq:1stmoment_CDE_a}) and $\epsilon$ times Eq.~(\ref{eq:2ndmoment_CDE_a}) by using $\partial_t=\partial_{t_0}+\epsilon\partial_{t_1}$, which  requires $\hat{n}^{(1)}_1$ and $\hat{n}^{(1)}_2$. These first order non-equilibrium moments ($\hat{n}^{(1)}_1$ and $\hat{n}^{(1)}_2$) can be obtained from Eqs.~(\ref{eq:1stmoment_CDE_b}) and (\ref{eq:1stmoment_CDE_c}), respectively, where the time derivatives are eliminated in favor of  the spatial derivatives by using the leading order mass, momentum and scalar conservation equations (i.e., Eqs.~(\ref{eq:1stmoment_fluidmotion_a}),~(\ref{eq:1stmoment_fluidmotion_b}),~(\ref{eq:1stmoment_fluidmotion_c}) and~(\ref{eq:1stmoment_CDE_a})). Hence after some simplification, and neglecting terms of $O(u^2)$ and higher, we get the components of the first order non-equilibrium moments in terms
of the components of the scalar gradient vector $\partial_i\phi$ as
\begin{subequations}
\begin{eqnarray}
\hat {n}^{(1)}_1=\hat{\eta}^{(1)'}_x&=&-\frac{1}{\omega_1^\phi}c_{s\phi}^2\partial_x\phi\label{eq:1stnoneqmmomentx_CDE}\\
\hat {n}^{(1)}_2=\hat{\eta}^{(1)'}_y&=&-\frac{1}{\omega_2^\phi}c_{s\phi}^2\partial_y\phi,\label{eq:1stnoneqmmomenty_CDE}
\end{eqnarray}
\end{subequations}
It may be noted that in the derivation of these non-equilibrium moment components, only the spatial derivatives of the second order moment equilibria (i.e., $\hat{\eta}_{xx}^{eq'}$, $\hat{\eta}_{yy}^{eq'}$ and $\hat{\eta}_{xy}^{eq'}$) are involved and do not involve the introduced anisotropy, which appears at a higher order level, i.e., for the third order moments of the equilibrium distribution via the factors $\beta_1$ and $\beta_2$ and hence the advection-diffusion of the passive scalar transport is correctly recovered.

As shown in the previous section, the symmetric components of the velocity gradient tensor $\partial_x u_x,\partial_y u_y$ and $\partial_x u_y+\partial_y u_x$ can be obtained from the MRT-LBM for fluid flow. In order to obtain the skew-symmetric component, i.e., $\partial_x u_y-\partial_y u_x$, which would then provide a complete information about the velocity gradient tensor $\partial_j u_i$ and hence the vorticity field, we now exploit the additional degree of freedom available in the off-diagonal, second-order non-equilibrium moment equation resulting from the MRT-LBM for CDE, i.e., Eq.~(\ref{eq:1stmoment_CDE_f}). Simplifying this equation by eliminating the time derivative in favor of spatial derivatives and eliminating higher order terms (i.e., $(O(u^2))$ and above), we get
\begin{equation}
\beta_1 c_{s\phi}^2 \partial_x(\phi u_y)+\beta_2 c_{s\phi}^2 \partial_y(\phi u_x)=-\omega_5^\phi \hat{n}^{(1)}_5,
\end{equation}
which can be rewritten as
\begin{equation}
\hat{n}^{(1)}_5=-\frac{c_{s\phi}^2}{\omega_5^\phi}\left[\phi(\beta_1\partial_xu_y+\beta_2\partial_y u_x)+(\beta_1u_y\partial_x \phi+\beta_2 u_x \partial_y \phi)\right].\label{eq:offdiagonal2ndnoneqmmoment_CDE}
\end{equation}
Clearly, the anisotropy introduced into the scalar flux components in the third order moment equilibria results in an additional flexibility via an independent equation given above (Eq.~(\ref{eq:offdiagonal2ndnoneqmmoment_CDE})). In  this equation, the gradients of the scalar field in the Cartesian coordinate directions $\partial_x \phi$ and $\partial_y \phi$ can be obtained locally from Eq.~(\ref{eq:1stnoneqmmomentx_CDE}) and Eq.~(\ref{eq:1stnoneqmmomenty_CDE}); and with the knowledge of the off-diagonal second-order non-equilibrium moment component $\hat{n}^{(1)}_5$, then Eq.~(\ref{eq:offdiagonal2ndnoneqmmoment_CDE}) represents an additional independent equation to compute the antisymmetric velocity gradient tensor component, which will be exploited further in the next section.

\section{\label{sec:vorticitycomputation} Derivation of local expressions for the complete velocity gradient tensor and vorticity field}
In order to independently determine the cross-derivative components of the velocity gradient tensor, i.e., $\partial_yu_x$ and $\partial_xu_y$, we combine the analysis presented in the two earlier sections. In particular, the Eq.~(\ref{eq:strainratexy}) resulting from the solution of the MRT-LBM for fluid flow and Eq.~(\ref{eq:offdiagonal2ndnoneqmmoment_CDE}) from the MRT-LBM for CDE, can be rewritten as
\begin{subequations}
\begin{eqnarray}
\partial_xu_y+\partial_yu_x&=&N_{xy},\label{eq:crossderivativerelation1}\\
\beta_1\partial_xu_y+\beta_2\partial_yu_x&=&N_{xy}^{\phi},\label{eq:crossderivativerelation2}
\end{eqnarray}
\end{subequations}
where, when $\phi \neq 0$,
\begin{subequations}
\begin{eqnarray}
N_{xy}&=&-\frac{\omega_5}{\rho c_{s}^2 }\hat m_5^{(1)},\label{eq:crossderivativeRHS1}\\
N_{xy}^{\phi}&=&-\frac{\omega_5^{\phi}}{\phi c_{s\phi}^2 }\hat n_5^{(1)}-\frac{1}{\phi}(\beta_1 u_y \partial_x \phi+\beta_2 u_x \partial_y \phi).\label{eq:crossderivativeRHS2}
\end{eqnarray}
\end{subequations}
Solving Eqs.~(\ref{eq:crossderivativerelation1}) and (\ref{eq:crossderivativerelation2}), we get following independent and local expressions for the off-diagonal components or the cross derivatives of the velocity field, which is one of the main results of this work:
\begin{subequations}
\begin{eqnarray}
\partial_xu_y&=&\frac{N_{xy}^{\phi}-\beta_2 N_{xy}}{\beta_1-\beta_2},\label{eq:offdiagvelderivativeresult1}\\
\partial_yu_x&=&\frac{\beta_1 N_{xy}- N_{xy}^{\phi}}{\beta_1-\beta_2}.\label{eq:offdiagvelderivativeresult2}
\end{eqnarray}
\end{subequations}
The diagonal components of the velocity gradient tensor, i.e., $\partial_xu_x$ and $\partial_yu_y$ follow from solving the Eqs.~(\ref{eq:strainratexxpyy}) and (\ref{eq:strainratexxmyy}) resulting from the MRT-LBE for the fluid motion, which reads as
\begin{subequations}
\begin{eqnarray}
\partial_xu_x&=&-\frac{1}{4c_s^2\rho}\left[\omega_3\hat m_3^{(1)}+\omega_4\hat m_4^{(1)}\right],\label{eq:diagvelderivativeresult1}\\
\partial_yu_y&=&-\frac{1}{4c_s^2\rho}\left[\omega_3\hat m_3^{(1)}-\omega_4\hat m_4^{(1)}\right],\label{eq:diagvelderivativeresult2}
\end{eqnarray}
\end{subequations}
and this completes the determination of all the components of the velocity gradient tensor. Finally, a local expression for the pseudo-vector, viz., the vorticity field $\bm{\omega}=\bm \nabla \times \bm u = (0,0,\omega_z)$ can be obtained by combining Eqs.~(\ref{eq:offdiagvelderivativeresult1}) and (\ref{eq:offdiagvelderivativeresult2}) as
\begin{equation}
\omega_z=\partial_x u_y-\partial_y u_x=\frac{2N_{xy}^{\phi}-(\beta_1+\beta_2)N_{xy}}{(\beta_1-\beta_2)},\label{eq:vorticityexpression}
\end{equation}
which is another key result arising from our analysis.

The terms $N_{xy}$ and $N_{xy}^{\phi}$ given in Eqs.~(\ref{eq:crossderivativeRHS1}) and (\ref{eq:crossderivativeRHS2}), respectively, which are needed in Eqs.~(\ref{eq:offdiagvelderivativeresult1}), (\ref{eq:offdiagvelderivativeresult2}) and (\ref{eq:vorticityexpression}) can be evaluated locally using
\begin{subequations}
\begin{eqnarray}
\hat{m}^{(1)}_5&=&{\widehat\kappa_{xy}^{'}}-{\widehat\kappa_{xy}^{eq'}}={\widehat\kappa_{xy}^{'}}-\rho u_x u_y,\\
\hat{n}^{(1)}_5&=&{\widehat\eta_{xy}^{'}}-{\widehat\eta_{xy}^{eq'}}={\widehat\eta_{xy}^{'}}-\phi u_x u_y,
\end{eqnarray}
\end{subequations}
and also since $\hat{n}^{(1)}_1={\widehat\eta_{x}^{'}}-{\widehat\eta_{x}^{eq'}}={\widehat\eta_{x}^{'}}-\phi u_x$ and $\hat{n}^{(1)}_2={\widehat\eta_{y}^{'}}-{\widehat\eta_{y}^{eq'}}={\widehat\eta_{y}^{'}}-\phi u_y$, and from Eqs.~(\ref{eq:1stnoneqmmomentx_CDE}) and (\ref{eq:1stnoneqmmomenty_CDE}), we have the required local expressions for the derivatives of the scalar field, which read as
\begin{equation}
\partial_x \phi=-\frac{\omega_{1}^{\phi}}{c_{s\phi}^2}[{\widehat\eta_{x}^{'}}-\phi u_x],\quad \partial_y, \phi=-\frac{\omega_{2}^{\phi}}{c_{s\phi}^2}[{\widehat\eta_{y}^{'}}-\phi u_y]
\end{equation}
Note that $\beta_1\approx 1$ and $\beta_2\approx 1$, but $\beta_1\neq \beta_2$ and are otherwise free parameters. We typically set $\beta_1=1, \beta_2=0.9$ in this work. In addition, the expressions for $\hat m_3^{(1)}$ and $\hat m_4^{(1)}$ needed in the diagonal components of the velocity gradient tensor, i.e., Eqs.~(\ref{eq:diagvelderivativeresult1}) and (\ref{eq:diagvelderivativeresult2}) can be written as
\begin{subequations}
\begin{eqnarray}
\hat m_3^{(1)}&=&({\widehat\kappa_{xx}^{'}}+{\widehat\kappa_{yy}^{'}})-(2c_s^2\rho+\rho(u_x^2+u_y^2)),\\
\hat m_4^{(1)}&=&({\widehat\kappa_{xx}^{'}}-{\widehat\kappa_{yy}^{'}})-\rho(u_x^2-u_y^2).
\end{eqnarray}
\end{subequations}
In the above, $\widehat\kappa_{xx}^{'}$, $\widehat\kappa_{yy}^{'}$, $\widehat\kappa_{xy}^{'}$, $\widehat\eta_{x}^{'}$, $\widehat\eta_{y}^{'}$ and $\widehat\eta_{xy}^{'}$ are the raw moment components of different orders of the respective distribution functions. The formulation presented above thus
allows local computation of the complete velocity gradient tensor and hence the vorticity field without relying on any finite difference approximations of the velocity field.

\section{\label{sec:resultsanddiscussion}Results and discussion}
In this section, we will perform a numerical validation study of the new DDF MRT-LB scheme for vorticity computation. In this regard, we will consider a set of well-defined benchmark flow problems for which analytical solutions or numerical results for the vorticity field are available or can be derived. In the simulations results presented in the following, the relaxation times for the second order moments of the MRT-LBM for the flow field ($\omega_4=\omega_5=1/\tau$) are chosen to specify the desired fluid viscosity, while those for the first order moments of the MRT-LBM for the scalar field ($\omega_1^\phi=\omega_2^\phi=1/\tau^\phi$) are prescribed to select the diffusivity. The relaxation times of all the higher order moments for both the LB schemes are set to unity for simplicity. Unless otherwise specified, we consider the use of lattice units, i.e., $\delta_x=\delta_t=1.0$ typical for LB simulations and a reference density of unity is considered in this work. For all the benchmark problems reported in what follows, we set the coefficients for the scalar flux terms in the third order moment equilibria of the MRT-LBM for the scalar field to $\beta_1=1.0$ and $\beta_2=0.9$.

\subsection{Poiseuille flow}
\begin{figure}[htbp]
\centering
\advance\leftskip-4cm
\advance\rightskip-4cm
        \includegraphics[scale=0.6] {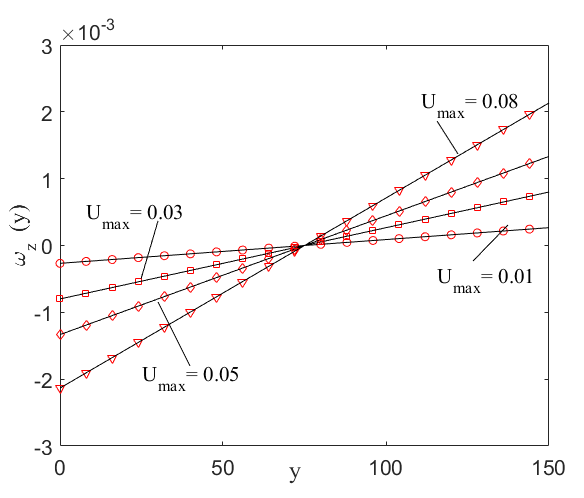}
        \caption{Comparison of the computed profiles of the vorticity field and the analytical solution in a Poiseuille flow for different values of the centerline velocity $U_{max}=0.01, 0.03, 0.05$, and $0.08$ obtained by varying the fluid viscosity at a fixed body force $F_x=3\times 10^{-6}$. Here, the lines represent the analytical solution and symbols refer to the numerical results obtained by the present DDF MRT-LB scheme.}
        \label{fig:Posieullevorticity}
\end{figure}

As the first benchmark problem, a steady flow between two parallel plates with a width $2L$ driven by a constant body force $F_x$, i.e., the Poiseuille flow, is simulated. This flow problem has an analytical solution for the vorticity field as the linear profile $\omega_z(y)=2U_{max}y/L ^2$, which can be obtained from the parabolic velocity profile $u_x(y)=U_{max}[1-\frac{y^2}{L^2}]$, where $U_{max}=\frac{F_xL^2}{2 \rho \nu}$ is the maximum centerline velocity,  $\nu$ and $\rho$ are fluid kinematic viscosity and density, respectively. Periodic boundary conditions are employed in the streamwise direction and no-slip condition for the velocity field are imposed using the half-way bounce back scheme. The computational domain is resolved using $3\times 151$ lattice nodes. For the scalar field, we consider fixed values at the bottom and top walls as $\phi_L=1.0$ and $\phi_H=2.0$, respectively, and its diffusivity is specified by choosing $\tau^\phi=0.57$. At a fixed body force $F_x=3\times 10^{-6}$, computations are carried by adjusting the fluid kinematic viscosity such that the following five sets of maximum centerline velocities are considered: $U_{max}=0.01,0.03,0.05$, and $0.08$.  Figure~\ref{fig:Posieullevorticity} shows a comparison between numerical results for the vorticity profiles obtained using the DDF MRT-LB scheme and the analytical solutions for the above set of values for $U_{max}$. Excellent agreement is seen.

\subsection{Four-rolls mill flow problem}
In order to examine the validity of our approach for a flow problem with fully two-dimensional (2D) spatially varying distribution of the vorticity field, we consider next the four-rolls mill flow. It is a steady, rotational flow consisting of an array of counter-rotating vortices generated by the stirring action of a suitably specified local body force $F_x=F_x(x,y)$ and $F_y=F_y(x,y)$ in a periodic square domain of size $2\pi \times 2\pi$. It is a modified form of the Taylor-Green vortex flow. The spatially varying driving body force can be written as $F_x(x,y)=2\rho_0\nu u_0 \sin x \sin y$ and $F_y(x,y) = 2\rho_0\nu u_0 \cos x \cos y$, where $\rho_0$ is the reference density, $\nu$  is kinematic viscosity and  $u_0 $ is the velocity scale and $0 \leq x,y \leq 2\pi$. A solution of the simplified form of the Navier-Stokes equations with the above described body force yields the explicit form of the local velocity field, which reads as $u_x(x,y)= u_0 \sin x \sin y$ and  $u_y(x,y) = u_0 \cos x \cos y$. Then, the analytical solution for the local vorticity field $\omega_z(x,y)$ can be derived by taking the curl of the above velocity field, which can be written as
\begin{eqnarray}
\omega_z(x,y)=-2u_0 \sin x\cos y.
\label{eq:mill}
\end{eqnarray}

\begin{figure}[htbp]
\centering
\advance\leftskip-3cm
\advance\rightskip-3cm
    \subfloat{
        \includegraphics[scale=0.45] {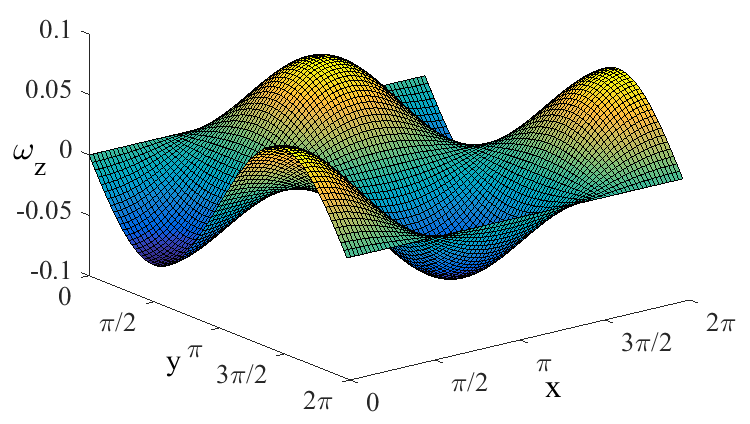}
        \label{fig:img12} } \hspace*{-22em}
    \hfill
    \subfloat{
        \includegraphics[scale=0.45] {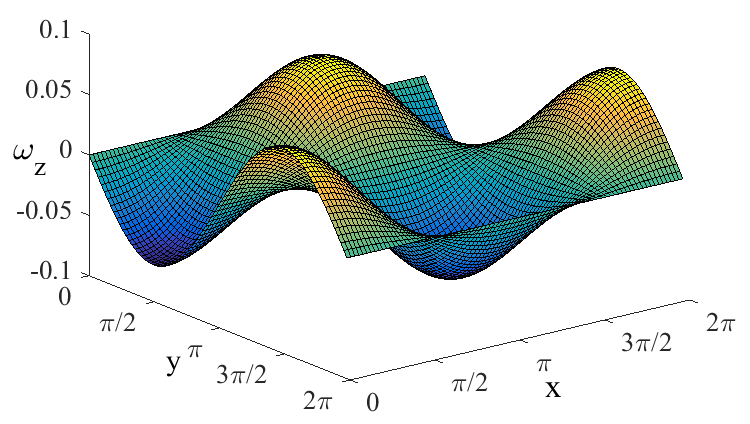}
        \label{fig:img24} } \\
    \caption{Comparison of the spatial distribution of the computed vorticity field with the analytical solution in a four-rolls mill flow within a square domain of size $2\pi \times 2\pi$ for $\mbox{Re}=40$. The surface plot on the left corresponds to the numerical results obtained by the present DDF MRT-LB scheme and that on the right is based on the analytical solution.}
    \label{fig:mill}
\end{figure}

\begin{figure}[htbp]
\centering
\advance\leftskip-1cm
\advance\rightskip-4cm
        \includegraphics[scale=0.6] {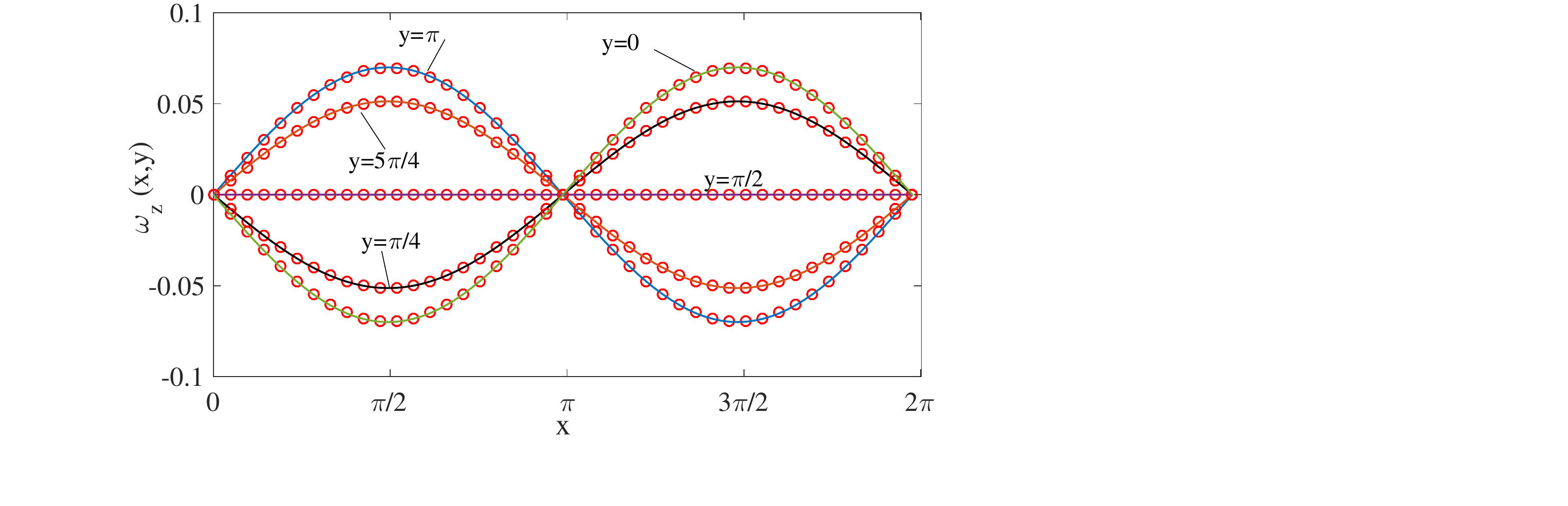}
        \caption{Comparison of computed profiles of the vorticity field and the analytical solution in a four-rolls mill flow along various horizontal sections at $y=0, \pi/4, \pi/2 ,\pi,5\pi/4$. Here, the lines represent the analytical solution and symbols refer to the numerical results obtained by the present DDF MRT-LB scheme.}
        \label{fig:vel}
\end{figure}

For the purpose of setting up simulations, the Reynolds number for this flow problem can be defined as $\mbox{Re}=u_02\pi/\nu$ and the viscosity can be written as $\nu=\frac{1}{3}(\tau-\frac{1}{2})\Delta x$, where $\Delta x=\Delta t=2\pi/N$, where $N$ is the number of grid nodes in each direction. We consider a grid resolution of $84\times 84$ and a velocity scale $u_0=0.035$ to simulate four-rolls mill flow at $\mbox{Re}=40$. The scalar field is initialized to a uniform value of $2.0$ in this periodic domain with the relaxation time $\tau^\phi=0.57$. Figure~\ref{fig:mill} presents a comparison between the spatial distribution of the computed vorticity field obtained using the DDF MRT-LB scheme and the analytical solution. Due to the
presence of a system of counter-rotating vortices, the vorticity field, represented by harmonic functions analytically, dramatically varies both in its magnitude and sign. Good agreement between the two results are evident.
Furthermore, in order to make a more head-on comparison, Fig.~\ref{fig:vel} shows the computed vorticity profiles $\omega_z(x,y)$ computed using our LB scheme along various horizontal sections at $y=0, \pi/4, \pi/2 ,\pi,5\pi/4$ along with results based on the analytical solution. It is evident that there is a very good agreement between our numerical results and the analytical solution.

\subsubsection{Grid convergence study}
\begin{figure}[htbp]
\centering
\advance\leftskip-1cm
\advance\rightskip-1cm
        \includegraphics[scale=0.75] {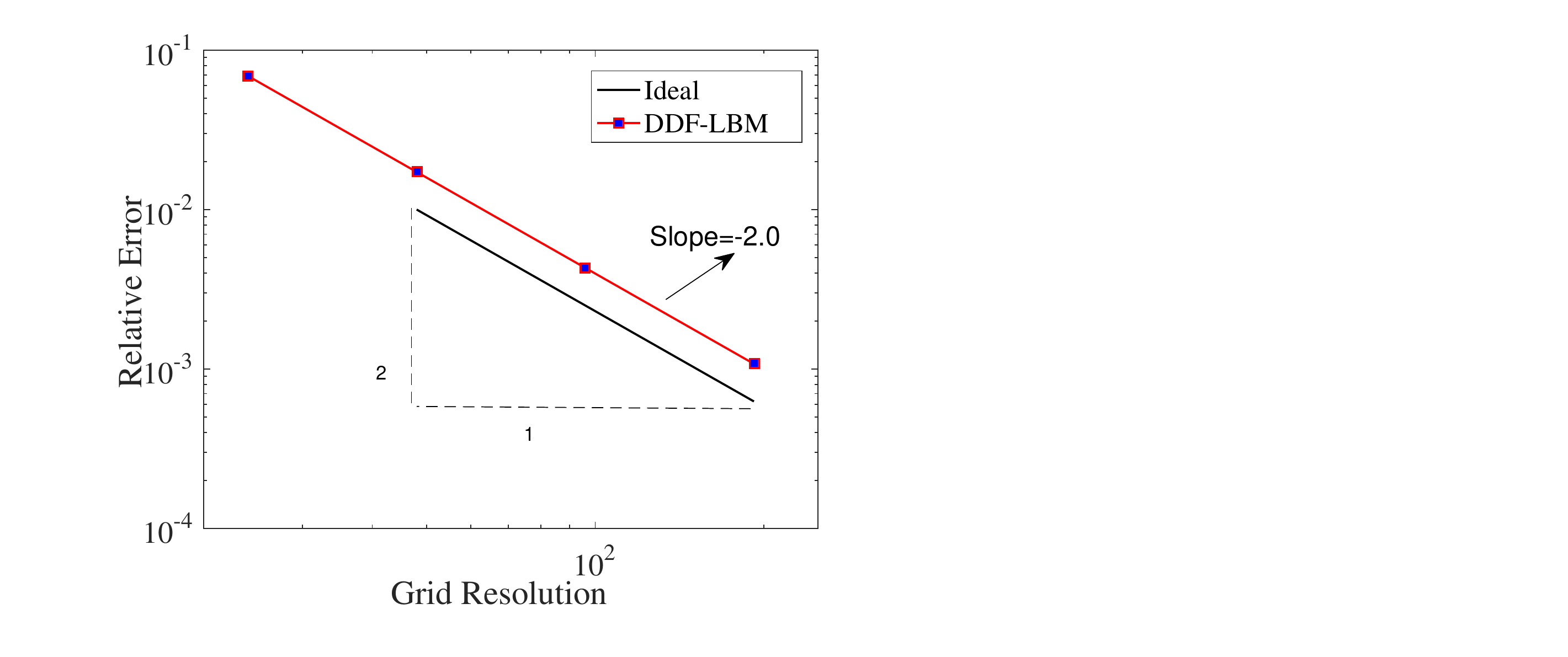}
        \caption{Evaluation of the order of accuracy of the present DDF MRT-LB scheme for vorticity computation in the four-rolls mill flow problem with a constant kinematic viscosity $\nu=0.00218$ at different grid resolutions.}
        \label{fig:Convergence}
\end{figure}

We will now assess the order of accuracy of the convergence of the vorticity computation via our DDF MRT-LB scheme. In this regard, at a fixed viscosity of $\nu=0.00218$ with a velocity scale $u_0=0.045$, we consider the following sequence of four different resolutions: $24\times 24$, $48\times 48$, $96\times 96$ and $192\times 192$. For each case, we measure the following global relative error ($E_{g,\omega}$) between the vorticity field computed using the DDF MRT-LB scheme given by $\omega_c$ and the corresponding analytical solution denoted by $\omega_a$:
\begin{eqnarray}
E_{g,\omega}=\sqrt{\frac{\Sigma (\omega_c-\omega_a)^2}{\Sigma{(\omega_a)^2}}},
\label{eq:sca}
\end{eqnarray}
where the summations in the above are for the whole computational domain. The rate of convergence of the global relative error is depicted using a log-log
scale in Fig.~\ref{fig:Convergence}. From this figure, it can  be seen that the relative error exhibits a slope of -2.0, which demonstrates that the vorticity computation using our approach is second order accurate.

\subsection{Womersley flow}
In order to validate our approach for the calculation of the vorticity field in unsteady flows, a 2D pulsatile flow between two parallel plates separated by a width $2L$ driven by a sinusoidally time-dependent body force $F_x(t)$ is considered. This classical Womersley flow problem is subjected to a periodic body force given by $F_x=F_m\mbox{cos}(\Omega t)$, where $F_m$ is the maximum amplitude of the force and $\Omega=2 \pi/T$ is the angular frequency and $T$ being the time period. Considering that this pulsatile flow is laminar and incompressible, the analytical solution for velocity field is given as~\cite{currie2002fundamental}
\begin{eqnarray}
&u(y,t)=\mathcal{R}\left\{i\frac{F_m}{\Omega}\right[1-\frac{\mbox {cos}(\gamma y/L)}{\mbox {cos}{\gamma}}]e^{i\Omega t} \},
\label{eq:wo}
\end{eqnarray}
where $\gamma=\sqrt{i\mbox{Wo}^2}$ and  $\mbox{Wo}= L \sqrt{(\Omega/\nu)}$  is the Womersley number. Here, and in the following $\mathcal{R}\{\cdot\}$ refers to the real part of the expression. Then, the analytical solution for the local time dependent vorticity field $\omega_z(y,t)$ can be readily obtained by taking the curl of the velocity field as
\begin{eqnarray}
&\omega_z(y,t)=\mathcal{R}\left\{i\frac{\gamma F_m}{\Omega L}\right[\frac{\mbox {sin}(\gamma y/L)}{\mbox {cos}{\gamma}}]e^{i\Omega t} \}.
\label{eq:wo}
\end{eqnarray}

\begin{figure}[htbp]
\centering
\advance\leftskip-1cm
\advance\rightskip1cm
    \subfloat{
     \includegraphics[trim=5cm 1cm 0cm 2cm,scale=0.45] {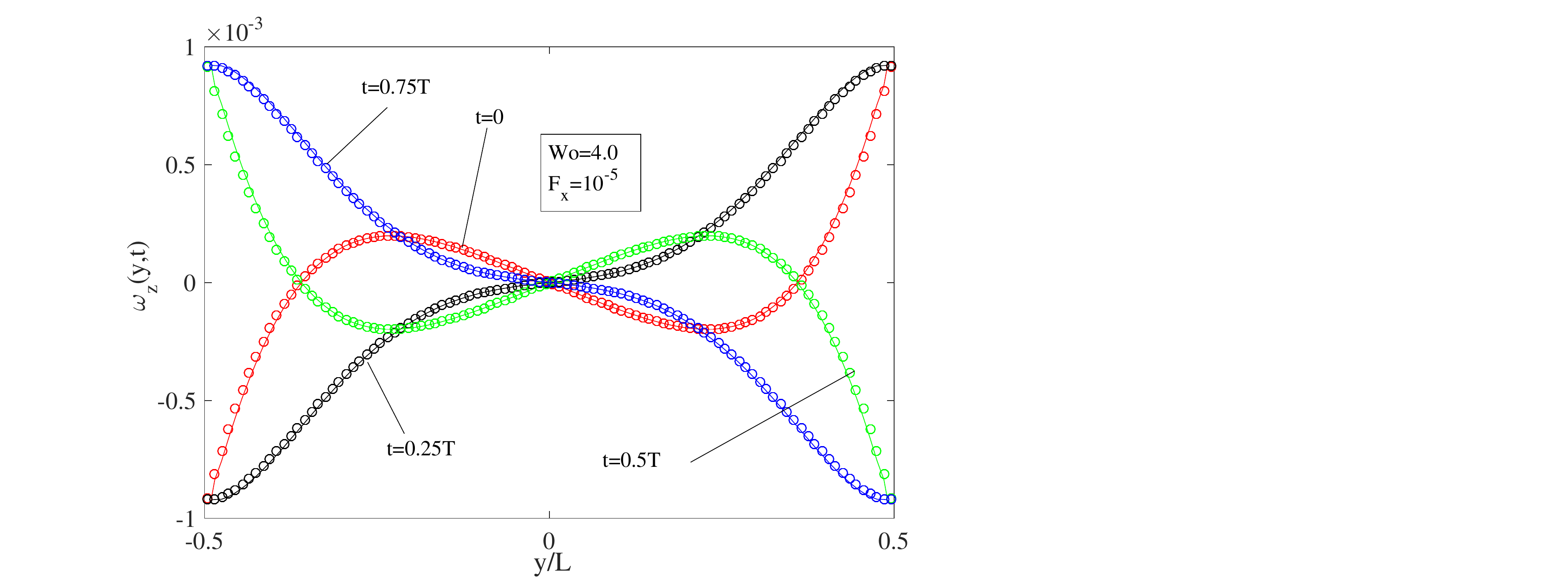}
        \label{fig:img12} } \hspace*{-20em}
    \hfill
    \subfloat{
     \includegraphics[trim=.5cm 1cm 2cm 3cm,scale=0.45]{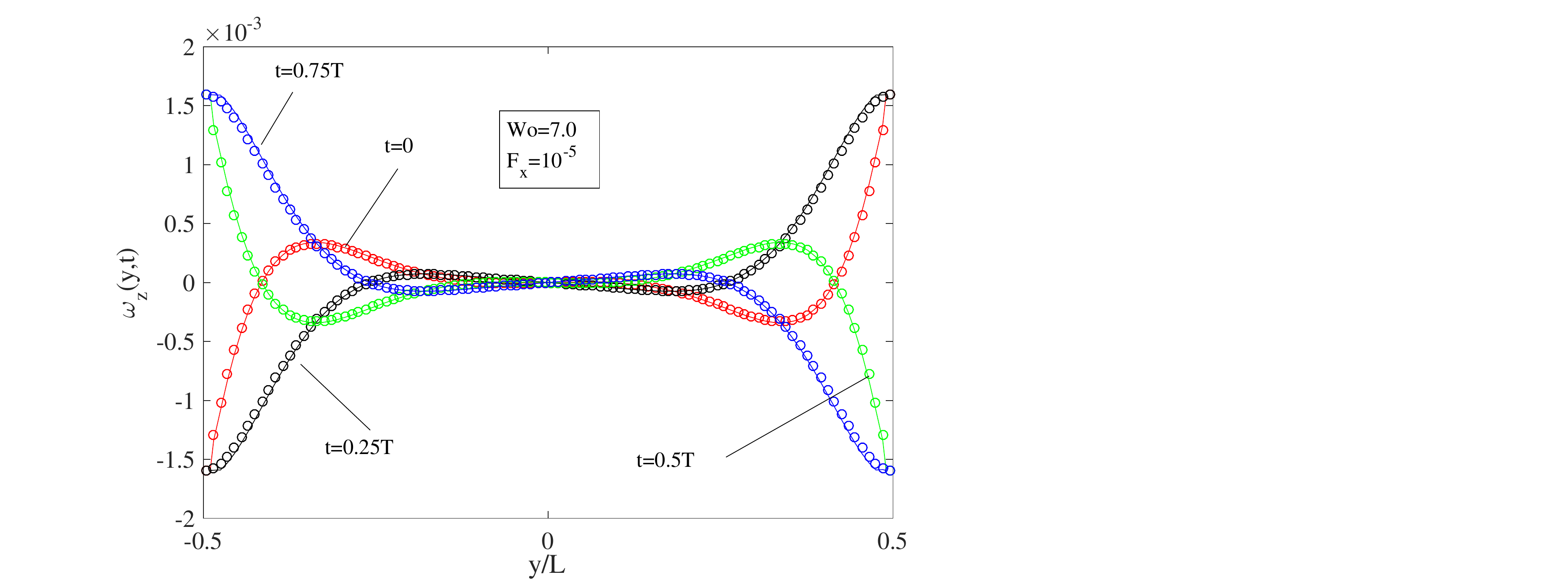}
        \label{fig:img24} } \\
    \caption{Comparison of computed profiles of the vorticity field and the analytical solution in a pulsatile flow in a channel (i.e., Womersley flow) at different instants within a time period for two different Womersley numbers of  $\mbox{Wo} = 4.0$ and $\mbox{Wo} = 7.0$. Here, lines represent the analytical solution and the symbols refer to the numerical results obtained using the present DDF MRT-LB scheme.}
    \label{fig:wom}
\end{figure}

We consider a grid resolution of $3\times 101$, maximum force amplitude $F_m=1.0\times 10^{-5}$ with a time period $T=10,000$ and two different values of the Womersley number, i.e., $\mbox{Wo}=4.0$ and $\mbox{Wo}=7.0$, which are specified by setting the relaxation times for the MRT-LBM for the flow field to be $\tau=0.781$ and $\tau=0.596$, respectively. Periodic boundary conditions and the no-slip boundary conditions are considered for the inlet/outlet in the streamwise direction and along the two parallel walls, respectively. The parameters and the boundary conditions for the scalar field are the same as those considered for the Poiseuille flow simulations discussed earlier. Figure~\ref{fig:wom} presents a comparison between the computed vorticity profiles obtained using the DDF MRT-LB scheme and the corresponding analytical solution at different time instants within a time period $T$. It is evident that the vorticity field is subjected to strong temporal and spatial variations, which are seen to increase with the Womersley number. These are very well reproduced quantitatively by our local computational approach.

\subsection{Lid-driven cavity flow}
\begin{figure}[htbp]
\centering
%\advance\leftskip-2cm
%\advance\rightskip3cm
    \subfloat{
        \includegraphics[scale=0.4,trim={0 3cm 0 0},clip] {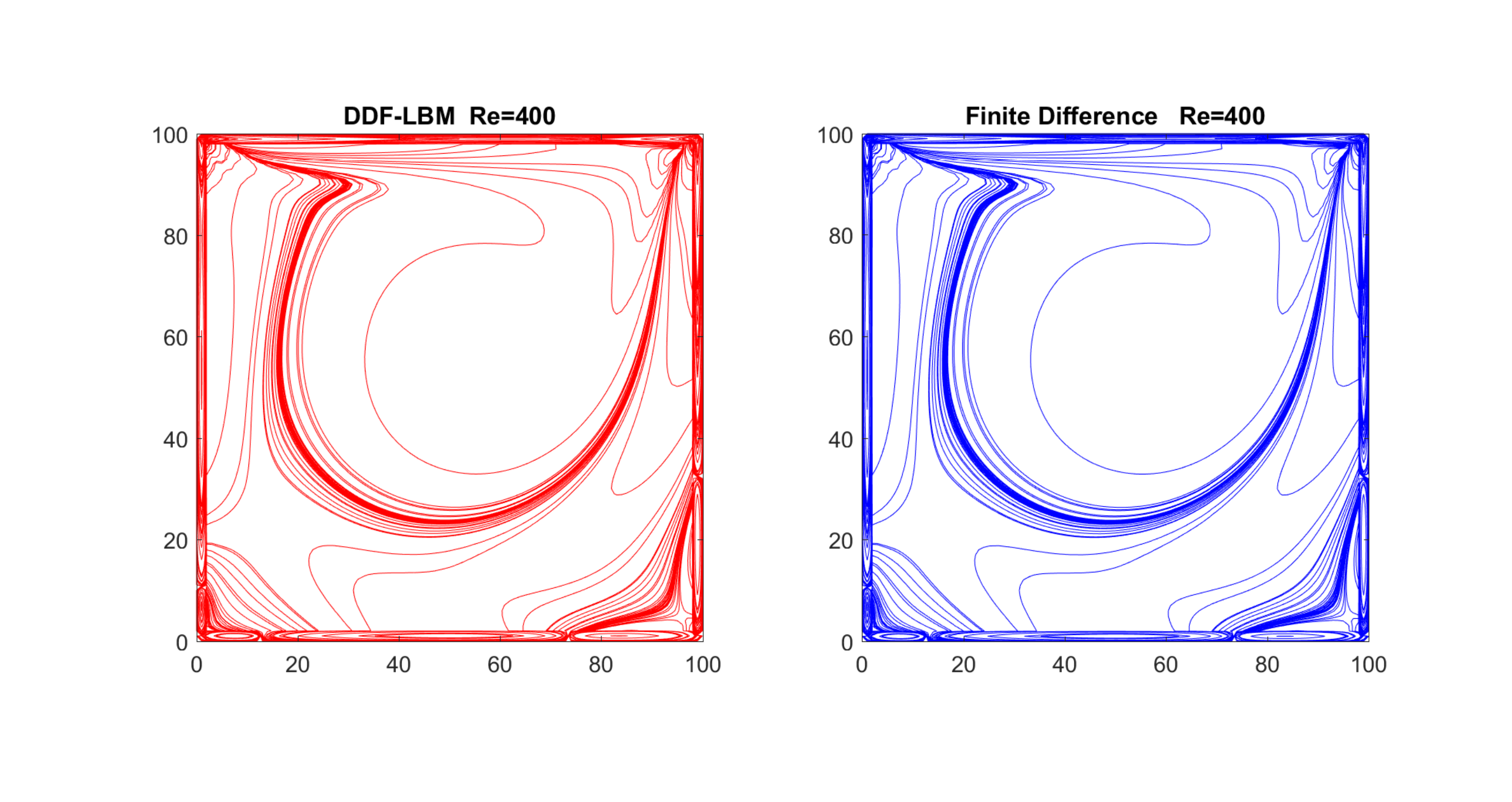}
         }
     \hfill
               \subfloat{
       \includegraphics[scale=0.4,trim={0 3cm 0 0},clip] {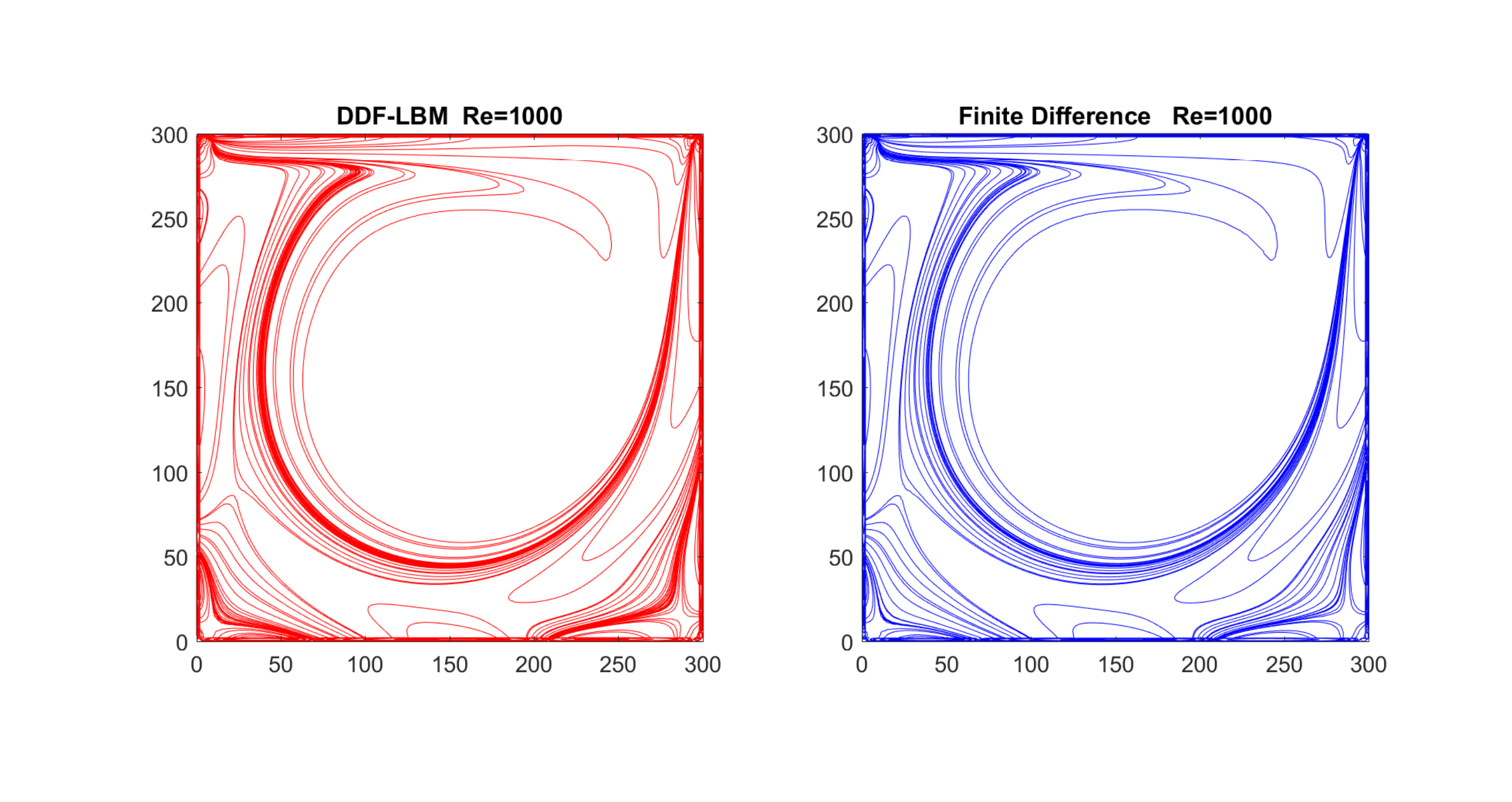}
        }
     \hfill
        % \advance\leftskip-2cm
          \subfloat{
        \includegraphics[scale=0.4,trim={0 3cm 0 0},clip] {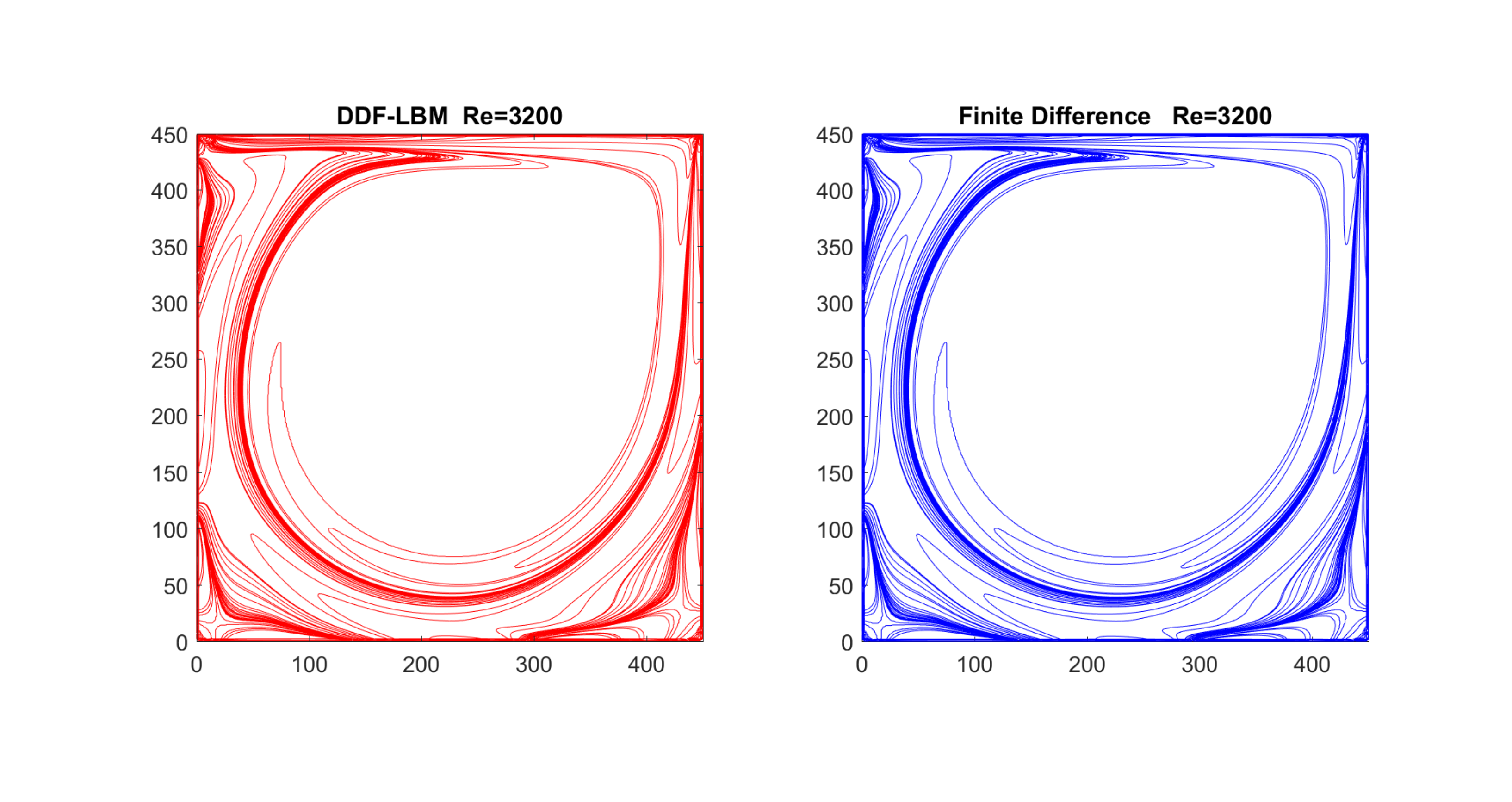}
         }
         \caption{: Comparison of computed contours of the vorticity field obtained using the DDF-LB scheme against the numerical solutions based on the finite-difference method for lid-driven cavity flow at three different Reynolds numbers: $\mbox{Re}=400$, $\mbox{Re}=1000$ and $\mbox{Re}=3200$.}
         \label{fig:liddrivencavityflow}
\end{figure}

As the final validation study, we consider simulation of a shear driven flow within a square cavity due to the motion of its top lid in order to compare the computed vorticity fields against those based on numerical results obtained by a finite-difference method. The lid-driven cavity flow is a classical benchmark problem characterized by complex flow features involving vortical patterns of different sizes which are strongly influenced by the nonlinear effects, i.e., the Reynolds number (see e.g.,~\cite{Ghia1982,erturk2005numerical,bruneau20062d}). If $U_0$ is the velocity imposed on the top lid of a square cavity of side length $L$, its Reynolds number $\mbox{Re}$ can be expressed as $\mbox{Re}=U_0L/\nu$. We perform numerical simulations of shear-driven flow within a cavity at $\mbox{Re} = 400, 1000$ and $3200$ by considering grid resolutions of $100\times 100$, $300\times 300$ and $450\times 450$. In this regard, the lid velocity $U_0$ is set to be $0.05$. The no-slip boundary conditions are prescribed on the walls via the standard half-way bounce-back condition, and including a momentum augmentation term for the moving top lid (see e.g.,~\cite{Ning2016} for details). The scalar field $\phi$ is set be equal to $1.0$ on all the boundaries. Simulations are carried out until steady state is reached in each case. Figure~\ref{fig:liddrivencavityflow} presents comparisons of the computed contours of the vorticity fields obtained using our DDF-LB scheme against numerical results based on the finite-difference
method at $\mbox{Re} = 400, 1000$ and $3200$.
The vorticity patterns are found to agree very well with one another. It is worth emphasizing that our DDF-LB
formulation used a completely local algorithm, while the finite-difference computations involved non-local operations to estimate the velocity derivatives. As $\mbox{Re}$ increases, the flow becomes progressively more complex involving the clustering of finer vortical features near walls, which are well reproduced by our scheme.

\section{\label{sec:summaryandconclusions}Summary and Conclusions}
A quantitative knowledge of the local skew-symmetric velocity gradient tensor, or equivalently the vorticity field, in conjunction with the
symmetric velocity gradient tensor is crucial for various applications, including those related to techniques for the identification of flow structures and in the modeling of complex fluids. In many situations, it is required to compute the fluid motion coupled to the transport by advection and diffusion of a scalar field. In the mesoscopic LB methods, the hydrodynamics (i.e., the NSE) and the scalar transport (i.e., the CDE) are commonly computed via the evolution of a pair of distribution functions represented by means of two LBMs. In such double distribution functions (DDF) based LB approaches, we present a new strategy for computing the vorticity field locally via exploiting the additional degrees of freedom available in the construction of the higher order moment equilibria in the collision model for the representation of the scalar transport to obtain the necessary additional independent relations. In particular, we have shown that this can be achieved by introducing an intensional anisotropy in the scalar flux components in the third order, off-diagonal moment equilibria, and then combining the second-order, off-diagonal non-equilibrium moment components of both the LB schemes. This approach for local vorticity computation has several advantages, which include the following. Any pair of lattice sets in the DDF-LBMs that support the third order off diagonal moments independently, which includes the various standard lattice velocity models in different dimensions, can allow a local determination of the complete flow kinematics, including the skew-symmetric velocity gradient tensor. It imposes no additional constraints on the higher order equilibrium moments of the LBM for the flow field, which can be solved by using any standard formulation without modification thereby
maintaining its numerical characteristics intact. Since the vorticity computation are based on distribution functions, which are generally solved to be second order accurate, the resulting mesoscopic and local computation of vorticity and the strain rate tensor are second order accurate as well. Moreover, the algorithm is completely local and do not depend on any finite difference approximations of the velocity derivatives, which is consistent with the general philosophy of the LBM as it well suited for implementation on parallel computers. The presented strategy is general and is applicable to a variety of collision models.

In the present work, for the purpose of demonstration, we formulate our approach by constructing in detail a DDF formulation using a MRT-LBM for the solution of the fluid motion and another MRT-LBM involving an anisotropy in the scalar flux components in the third order equilibria for the transport of a scalar field, each on a D2Q9 lattice.  By means of a Chapman-Enskog analysis, we have shown that the former provides the necessary second order non-equilibrium moment equations to determine the symmetric velocity gradient tensor, while the latter yields additional corresponding moment relations to obtain the skew-symmetric velocity gradient tensor. For simplicity, the MRT-LBMs are constructed using natural, non-orthogonal moment bases. In order to validate our new approach, we have presented comparisons of the computed vorticity fields against the analytical and/or solutions for various benchmark problems such as the steady flow in a channel, four-rolls mill flow, time-dependent pulsatile (Womersley) flow in a channel, and lid-driven cavity flow at different Reynolds numbers, which demonstrate its good accuracy. In addition, an analysis of the method for various grid resolutions establishes its second order convergence for computing vorticity. While the focus of this work is on presenting and validating a new method for local computation of vorticity field using a DDF-LB scheme based on a MRT formulation, we have also discussed its deveopment for other collision models such as those based on SRT and cascaded central moments. Extensions our approach to 3D for various standard lattice sets (i.e., D3Q15, D3Q19 and D3Q27) and using a variety of collision models~\cite{vorticity3Dtechreport} will be reported in a future investigation. It may be noted that the method presented here can be extended to vectorial~\cite{dellar2005lattice} and tensorial~\cite{denniston2001lattice} forms of distribution functions to model and locally compute the skew-symmetric velocity gradient contributions in the general constitutive relations for complex fluids. Moreover, spin relaxation to the vorticity and the coupling of the intrinsic angular momentum to the linear momentum need to be accounted for in molecular liquid flows in nanoscale confined geometries~\cite{hansen2009molecular,hansen2011nanoflow,hansen2015continuum}, which can be modeled as generalization of the Cosserat theory for micropolar fluids~\cite{dahler1963theory,eringen1966theory,eringen1964simple,de2013non}. The approach presented here can also be used to construct LB schemes to locally represent such effects, which are subjects for future studies.

\section*{Acknowledgements}
The authors would like to acknowledge the support of the US National Science Foundation (NSF) under Grant CBET-1705630.

\appendix
\section{\label{app:relation_noneqmoments_spatial_derivatives_eqm_moments}Relation between non-equilibrium moments and spatial derivatives of components of moment equilibria for D2Q9 lattice}
For better clarity, the $O(\epsilon)$ moment system using a non-orthogonal moment basis given in Eq.~(\ref{eq:CEmoment1stfluidmotion}) in Sec.~\ref{sec:LBEfluidmotion}, i.e., $(\partial_{t_0}+\widehat{\tensor E}_i \partial_i)\mathbf{\widehat{m}}^{(0)}=-\widehat{\tensor\Lambda}\mathbf{\widehat{m}}^{(1)}+\mathbf{\widehat{S}}$, which forms a main element in the derivation, can be expanded explicitly in terms of their various components as follows:
\begin{subequations}
\begin{eqnarray}
&\partial_{t_0}\hat{\kappa}^{eq'}_{0}+\partial_x \hat{\kappa}^{eq'}_{x}+\partial_y \hat{\kappa}^{eq'}_{y} = \hat{\sigma}^{'}_{0},
\nonumber\\
&\partial_{t_0}\hat{\kappa}^{eq'}_{x}+\partial_x \hat{\kappa}^{eq'}_{xx}+\partial_y \hat{\kappa}^{eq'}_{xy} = \hat{\sigma}^{'}_{x},
\nonumber\\
&\partial_{t_0}\hat{\kappa}^{eq'}_{y}+\partial_x \hat{\kappa}^{eq'}_{xy}+\partial_y \hat{\kappa}^{eq'}_{yy} = \hat{\sigma}^{'}_{y},
\nonumber\\
&\partial_{t_0}(\hat{\kappa}^{eq'}_{xx}+\hat{\kappa}^{eq'}_{yy})+\partial_x (\hat{\kappa}^{eq'}_{x}+\hat{\kappa}^{eq'}_{xyy})+\partial_y (\hat{\kappa}^{eq'}_{y}+\hat{\kappa}^{eq'}_{xxy}) = -\omega_3\hat{m}^{(1)}_3+\hat{\sigma}^{'}_{xx}+\hat{\sigma}^{'}_{yy},
\nonumber\\
&\partial_{t_0}(\hat{\kappa}^{eq'}_{xx}-\hat{\kappa}^{eq'}_{yy})+\partial_x (\hat{\kappa}^{eq'}_{x}-\hat{\kappa}^{eq'}_{xyy})+\partial_y (-\hat{\kappa}^{eq'}_{y}+\hat{\kappa}^{eq'}_{xxy}) = -\omega_4\hat{m}^{(1)}_4+\hat{\sigma}^{'}_{xx}-\hat{\sigma}^{'}_{yy},
\nonumber\\
&\partial_{t_0}\hat{\kappa}^{eq'}_{xy}+\partial_x \hat{\kappa}^{eq'}_{xxy}+\partial_y \hat{\kappa}^{eq'}_{xyy} = -\omega_5\hat{m}^{(1)}_5+\hat{\sigma}^{'}_{xy},
\nonumber\\
&\partial_{t_0}\hat{\kappa}^{eq'}_{xxy}+\partial_x \hat{\kappa}^{eq'}_{xy}+\partial_y \hat{\kappa}^{eq'}_{xxyy} = -\omega_6\hat{m}^{(1)}_6+\hat{\sigma}^{'}_{xxy},
\nonumber\\
&\partial_{t_0}\hat{\kappa}^{eq'}_{xyy}+\partial_x \hat{\kappa}^{eq'}_{xxyy}+\partial_y \hat{\kappa}^{eq'}_{xy} = -\omega_7\hat{m}^{(1)}_7+\hat{\sigma}^{'}_{xyy},
\nonumber\\
&\partial_{t_0}\hat{\kappa}^{eq'}_{xxyy}+\partial_x \hat{\kappa}^{eq'}_{xyy}+\partial_y \hat{\kappa}^{eq'}_{xyy} = -\omega_8\hat{m}^{(1)}_8+\hat{\sigma}^{'}_{xxyy}.
\nonumber
\end{eqnarray}
\end{subequations}
In general, it can be seen that any non-equilibrium moment of order $n$ depends on the spatial derivatives of equilibrium moments of order $(n+1)$ and $(n-1)$. In particular, the diagonal components of the second order moment ($\hat{m}^{(1)}_3$ and $\hat{m}^{(1)}_4$) depend on the moment equilibria of
first order ($\hat{\kappa}^{eq'}_{x}$ and $\hat{\kappa}^{eq'}_{y}$) and third order ($\hat{\kappa}^{eq'}_{xxy}$ and $\hat{\kappa}^{eq'}_{xyy}$),  while the off-diagonal second order moment ($\hat{m}^{(1)}_5$) depends only on those of the third order equilibrium moments ($\hat{\kappa}^{eq'}_{xxy}$ and $\hat{\kappa}^{eq'}_{xyy}$). These considerations are important in establishing the relationship between the non-equilibrium second-order moments and the velocity gradient tensor components. In the case of the LBE for computing fluid flow, the symmetry of their moment equilibria to respect the isotropy of the viscous stress tensor limits the dependence of the corresponding non-equilibrium second order moments to only on the symmetric part of the velocity gradient tensor (i.e., the strain rate tensor). However, the construction of the LBE for computing the transport of a passive scalar represented by the CDE does not need to satisfy these restrictive constraints, and the additional degrees of freedom available for the higher order moments can be suitably exploited~\cite{Hajabdollahiphdthesis}. Indeed, since the diffusion term of the CDE need only to satisfy a lower degree of isotropy than that of the viscous term of the NSE, the third order moment equilibria for solving the former case can be specifically designed to locally represent the skew-symmetric part of the velocity gradient tensor via the respective off-diagonal non-equilibrium second-order moment (based on an equation analogous to the sixth equation in the above moment system with ${\hat\kappa_{x^my^n}^{eq'}}$ replaced by ${\hat\eta_{x^my^n}^{eq'}}$ and $\hat {m}^{(1)}_j$ by $\hat {n}^{(1)}_j$ -- Sec.~\ref{sec:LBEscalartransport}).

\section{\label{app:SRT_LBM_CDE}SRT-LBM for solution of scalar transport to recover the skew-symmetric velocity gradient tensor}
In this appendix, we will present a special case of the single-relaxation-time (SRT)-LBM for the solution of the convection-diffusion equation of a passive scalar field to recover the skew-symmetric velocity gradient tensor. This can be written as
\begin{subequations}
\begin{eqnarray}
\tilde{g}_\alpha(\bm{x},t)&=& g_\alpha(\bm{x},t)-\frac{1}{\tau_\phi}[g_\alpha(\bm{x},t)-g_\alpha^{eq}(\bm{x},t)],\label{eq:SRT_LBM_CDEa}\\
g_\alpha(\bm{x},t+\delta_t)&=& \tilde{g}_\alpha(\bm{x}-\bm{e}_\alpha\delta_t,t),\label{eq:SRT_LBM_CDEb}
\end{eqnarray}
\end{subequations}
where the post-collision distribution functions $\tilde{g}_\alpha$ are prescribed by an update of $g_\alpha$ involving their relaxation to local equilibrium distribution functions $g_\alpha^{eq}$ at a single relaxation time $\tau_\phi$. A key aspect here is the construction of $g_\alpha^{eq}=g_\alpha^{eq}(\phi,\bm{u},\beta_1,\beta_2)$ that facilitates the recovery of $\Omega_{ij}=\frac{1}{2}(\partial_j u_i - \partial_i u_j)$, which we achieve by mapping the various equilibrium moment components derived earlier for the D2Q9 lattice (see Sec.~\ref{sec:LBEscalartransport}) to the velocity space. In this regard, defining
\begin{eqnarray}
\hat {\mathbf {q}}^{eq}&=&(\hat {q}_0^{eq},\hat {q}_1^{eq},\hat {q}_2^{eq}\dots \hat {q}_8^{eq})^\dagger\nonumber\\
&=&(\hat{\eta}_0^{eq^\prime},\hat{\eta}_x^{eq^\prime},\hat{\eta}_y^{eq^\prime},\hat{\eta}_{xx}^{eq^\prime},\hat{\eta}_{yy}^{eq^\prime},\hat{\eta}_{xy}^{eq^\prime},\hat{\eta}_{xxy}^{eq^\prime},\hat{\eta}_{xyy}^{eq^\prime},\hat{\eta}_{xxyy}^{eq^\prime})^\dagger,\label{eq:momentequilibriaCDE1}
\end{eqnarray}
we can relate it to ${\mathbf {g}}^{eq}=(g_0^{eq},g_1^{eq},g_2^{eq}\dots g_8^{eq})^\dagger$ via $\hat {\mathbf {q}}^{eq}=\tensor P \mathbf {g}^{eq}$ using the bare moment basis $\tensor P$ given by
\begin{equation}
\tensor{P}=\left[{{P}_{0}},{{P}_{1}},{{P}_{2}},{{P}_{3}},{{P}_{4}},{{P}_{5}},{{P}_{6}},{{P}_{7}},{{P}_{8}}\right],
\label{eq:transformationmatrix1}
\end{equation}
where
\begin{eqnarray}
&&{{P}_{0}}=\ket{1}, \quad{{P}_{1}}=\ket{e_{ x}},\quad{{P}_{2}}=\ket{e_{ y}}, \quad {{P}_{3}}=\ket{e_{ x}^2}, \quad {{P}_{4}}= \ket{e_{ y}^2}, \nonumber \\
&&{{P}_{5}}=\ket{e_{ x}e_{ y}},\quad {{P}_{6}}=\ket{e_{ x}^2e_{ y}}, \quad {{P}_{7}}=\ket{e_{ x}e_{ y}^2},\quad {{P}_{8}}=\ket{e_{ x}^2e_{ y}^2}.\label{eq:basisvectors1}
\end{eqnarray}
Inverting, that is,
\begin{eqnarray}
\mathbf {g}^{eq} =\tensor P^{-1}  \hat {\mathbf {q}}^{eq},
\end{eqnarray}
we can then obtain the one-to-one mapping between the equilibrium moments and the equilibrium distribution functions. Thus, we have
\begin{subequations}
\begin{eqnarray}
g_0^{eq} &=& \hat{\eta}_0^{eq^\prime} - \hat{\eta}_{xx}^{eq^\prime} - \hat{\eta}_{yy}^{eq^\prime} + \hat{\eta}_{xxyy}^{eq^\prime},\\
g_1^{eq} &=& \frac{1}{2}\left[\hat{\eta}_{x}^{eq^\prime}+\hat{\eta}_{xx}^{eq^\prime}-\hat{\eta}_{xyy}^{eq^\prime}-\hat{\eta}_{xxyy}^{eq^\prime}\right],\\
g_2^{eq} &=& \frac{1}{2}\left[\hat{\eta}_{y}^{eq^\prime}+\hat{\eta}_{yy}^{eq^\prime}-\hat{\eta}_{xxy}^{eq^\prime}-\hat{\eta}_{xxyy}^{eq^\prime}\right],\\
g_3^{eq} &=& \frac{1}{2}\left[-\hat{\eta}_{x}^{eq^\prime}+\hat{\eta}_{xx}^{eq^\prime}+\hat{\eta}_{xyy}^{eq^\prime}-\hat{\eta}_{xxyy}^{eq^\prime}\right],\\
g_4^{eq} &=& \frac{1}{2}\left[-\hat{\eta}_{y}^{eq^\prime}+\hat{\eta}_{yy}^{eq^\prime}+\hat{\eta}_{xxy}^{eq^\prime}-\hat{\eta}_{xxyy}^{eq^\prime}\right],\\
g_5^{eq} &=& \frac{1}{4}\left[\hat{\eta}_{xy}^{eq^\prime}+\hat{\eta}_{xxy}^{eq^\prime}+\hat{\eta}_{xyy}^{eq^\prime}+\hat{\eta}_{xxyy}^{eq^\prime}\right],\\
g_6^{eq} &=& \frac{1}{4}\left[-\hat{\eta}_{xy}^{eq^\prime}+\hat{\eta}_{xxy}^{eq^\prime}-\hat{\eta}_{xyy}^{eq^\prime}+\hat{\eta}_{xxyy}^{eq^\prime}\right],\\
g_7^{eq} &=& \frac{1}{4}\left[\hat{\eta}_{xy}^{eq^\prime}-\hat{\eta}_{xxy}^{eq^\prime}-\hat{\eta}_{xyy}^{eq^\prime}+\hat{\eta}_{xxyy}^{eq^\prime}\right],\\
g_8^{eq} &=& \frac{1}{4}\left[-\hat{\eta}_{xy}^{eq^\prime}-\hat{\eta}_{xxy}^{eq^\prime}+\hat{\eta}_{xyy}^{eq^\prime}+\hat{\eta}_{xxyy}^{eq^\prime}\right],
\end{eqnarray}
\end{subequations}
where $\hat{\eta}_{x^my^n}^{eq^\prime}$ are given in Eq.~(\ref{eq:eqmrawmoment_CDE}). In particular, the third order moment equilibrium components $\hat{\eta}_{xxy}^{eq^\prime}$ and $\hat{\eta}_{xyy}^{eq^\prime}$ contain the intensional anisotropy needed for recovering the skew-symmetric velocity gradient tensor. Setting the relaxation time $\tau_\phi$ in terms of the relaxation parameter $\omega_j^\phi$ as $\omega_j^\phi=1/\tau_\phi$ and using the
definitions of $\hat{\eta}_{x^my^n}^{^\prime}$ and $\hat{\eta}_{x^my^n}^{eq^\prime}$ given in Sec.~\ref{sec:LBEscalartransport}, the local expressions derived earlier in Sec.~\ref{sec:vorticitycomputation} for $N_{xy}^\phi$, $\hat n_5^{(1)}$, $\partial_x \phi$, $\partial_y \phi$, $\partial_x u_y$ and $\partial_y u_x$ and $\omega_z$ in terms of the non-equilibrium moments are valid.

\section{\label{app:Cascaded_LBM_CDE}Cascaded LBM based on central moments for solution of scalar transport to recover the skew-symmetric velocity gradient tensor}
In this section, we will present further development of our formulation to a more general cascaded LBM based on central moments~\cite{Geier2006} extended for the solution of a scalar transport~\cite{Hajabdollahi2018,hajabdollahi2018symmetrized,hajabdollahi2019cascaded} capable of locally computing the skew-symmetric velocity gradient tensor. In this regard, we define the central moments of the distribution functions and their equilibrium as
\begin{eqnarray}
\left( \begin{array}{l}
{{\hat \eta }_{{x^m}{y^n}}}\\
{\hat \eta}^{eq}_{{x^m}{y^n}}
\end{array} \right) = \sum\limits_{\alpha=0}^8  {\left( \begin{array}{l}
{g_\alpha }\\
g_\alpha ^{eq}\\
\end{array} \right)}{(e_{\alpha x}-u_x)^m}{(e_{\alpha y}-u_y)^n}.\label{eq:centralmomentdefinitions_CDE}
\end{eqnarray}
We prescribe the central moment equilibria based on those of the local Maxwellian, by replacing the density with the scalar field $\phi$ (see e.g.,~\cite{Hajabdollahi2018,hajabdollahi2018symmetrized,hajabdollahi2019cascaded}). Usually, the third order central moment equilibria then
become $\hat \eta^{eq}_{xxy}=\hat \eta^{eq}_{xyy} = 0$ and the corresponding raw moment equilibria are $\hat \eta^{eq\prime}_{xxy} = c_{s\phi}^2\phi u_y+\phi u_x^2u_y$ and $\hat \eta^{eq\prime}_{xyy} = c_{s\phi}^2\phi u_x+\phi u_xu_y^2$~\cite{Hajabdollahi2018,hajabdollahi2018symmetrized,hajabdollahi2019cascaded}. On the other hand, to enable local computation of the vorticity
field, our derivation in Secs.~\ref{sec:LBEscalartransport} and~\ref{sec:vorticitycomputation} required the above raw moment components to be modified to
$\hat \eta^{eq\prime}_{xxy} = \beta_1c_{s\phi}^2\phi u_y+\phi u_x^2u_y$ and $\hat \eta^{eq\prime}_{xyy} = \beta_2c_{s\phi}^2\phi u_x+\phi u_xu_y^2$. These are equivalent to modifying the central moment equilibria $\hat \eta^{eq}_{xxy}$ and $\hat \eta^{eq}_{xxy}$ as $\hat \eta^{eq}_{xxy} = (\beta_1-1)c_{s\phi}^2\phi u_y$ and $\hat \eta^{eq}_{xyy} = (\beta_2-1)c_{s\phi}^2\phi u_x$, where $(\beta_1-1)$ and $(\beta_2-1)$ represent the degree of anisotropy in the scalar flux components $\phi u_y$ and $\phi u_x$, respectively. Hence, we enumerate all the central moment equilibria for the D2Q9 lattice as
\begin{eqnarray}
&\widehat{\eta}^{eq}_{0}=\phi,\quad
\widehat{\eta}^{eq}_{x}=\widehat{\eta}^{eq}_{y}=0, \quad
\widehat{\eta}^{eq}_{xx}=\widehat{\eta}^{eq}_{yy}=c_{s\phi}^{2}\phi,\quad
\widehat{\eta}^{eq}_{xy}=0, \nonumber\\
&\widehat{\eta}^{eq}_{xxy}=\boxed{(\beta_1-1)c_{s\phi}^{2}\phi u_y},\quad
\widehat{\eta}^{eq}_{xyy}=\boxed{(\beta_2-1)c_{s\phi}^{2}\phi u_x},\quad
\widehat{\eta}^{eq}_{xxyy}=c_{s\phi}^{4}\phi.\label{eq:eqmcentralmoment_CDE}
\end{eqnarray}

The cascaded LBM then reads as
\begin{subequations}
\begin{eqnarray}
\tilde{g}_\alpha(\bm{x},t)&=& g_\alpha(\bm{x},t)+(\tensor{K}\cdot\widehat{\mathbf{h}})_\alpha,\label{eq:cascaded_LBMa}\\
g_\alpha(\bm{x},t+\delta_t)&=& \tilde{g}_\alpha(\bm{x}-\bm{e}_\alpha\delta_t,t),\label{eq:cascaded_LBMb}
\end{eqnarray}
\end{subequations}
where $\widehat{\mathbf{h}}=(\widehat{h}_0,\widehat{h}_1,\cdots,\widehat{h}_8)^\dag$ represents the changes in different moments due to collision via
relaxation of central moments in a cascaded fashion. Here, $\tensor{K} = (K_0,K_1,\cdots,K_8)^\dag$ represents a matrix holding the orthogonal basis vectors given by
\begin{eqnarray}
&&{{K}_{0}}=\ket{1}, \quad{{K}_{1}}=\ket{e_{ x}},\quad{{K}_{2}}=\ket{e_{ y}}, \quad {{K}_{3}}=3\ket{e_{ x}^2+e_{ y}^2}-4\ket{1}, \nonumber \\ &&{{K}_{4}}=\ket{e_{ x}^2-e_{ y}^2}, \quad{{K}_{5}}=\ket{e_{ x}e_{ y}},\quad {{K}_{6}}=\ket{e_{ x}^2e_{ y}}, \quad {{K}_{7}}=\ket{e_{ x}e_{ y}^2},\nonumber\\
&&{{K}_{8}}=9\ket{e_{ x}^2e_{ y}^2}-6\ket{e_{ x}^2+e_{ y}^2}+4\ket{1}.\label{eq:orthogonalbasisvectors}
\end{eqnarray}
To obtain the change in moments under collision $\widehat{\mathbf{h}}$, we need the following inner products:
\begin{eqnarray}
&&\braket{1|\tensor{K}\cdot\widehat{\mathbf{h}}} = 0, \quad\braket{e_{x}|\tensor{K}\cdot\widehat{\mathbf{h}}} = 6\widehat{h}_1, \quad\braket{e_{y}|\tensor{K}\cdot\widehat{\mathbf{h}}} = 6\widehat{h}_2,\nonumber\\
&&\braket{e_{x}^2|\tensor{K}\cdot\widehat{\mathbf{h}}} = 6\widehat{h}_3+2\widehat{h}_4,\quad\braket{e_{y}^2|\tensor{K}\cdot\widehat{\mathbf{h}}} = 6\widehat{h}_3-2\widehat{h}_4,\quad\braket{e_{x}e_{y}|\tensor{K}\cdot\widehat{\mathbf{h}}} = 4\widehat{h}_5,\nonumber\\
&&\braket{e_{x}^2e_{y}|\tensor{K}\cdot\widehat{\mathbf{h}}} = 4\widehat{h}_2-4\widehat{h}_6,\quad\braket{e_{x}e_{y}^2|\tensor{K}\cdot\widehat{\mathbf{h}}} = 4\widehat{h}_1-4\widehat{h}_7,\nonumber\\
&&\braket{e_{x}^2e_{y}^2|\tensor{K}\cdot\widehat{\mathbf{h}}} = 8\widehat{h}_3+4\widehat{h}_8.\label{eq:innerproductscollisionchanges}
\end{eqnarray}
Then, we prescribe the relaxation of various central moments to their corresponding equilibria supported by the D2Q9 lattice as
\begin{equation}
\braket{(\mathbf{e_x}-u_x\mathbf{1})^m(\mathbf{e_y}-u_y\mathbf{1})^n|\tensor{K}\cdot\widehat{\mathbf{h}}} = \omega_*^\phi[\hat{\eta}^{eq}_{x^my^n}-\hat{\eta}_{x^my^n}], \label{eq:centralmomentrelaxcationcascaded}
\end{equation}
where $\mathbf{1}=\ket{1}$, $\mathbf{e}_x=\ket{e_x}$, $\mathbf{e}_y=\ket{e_y}$ and $\omega_*^\phi$ being the relaxation parameter of the central moment
of order $(m+n)$. With the zeroth moment being conserved, i.e., a collision invariant, and evaluating Eq.~(\ref{eq:centralmomentrelaxcationcascaded}) at various orders and then simplifying the resulting expressions, we obtain the changes in different moments due to cascaded collision as
\begin{eqnarray}
\widehat{h}_0&=&0,\nonumber\\
\widehat{h}_1&=&\frac{\omega_1^\phi}{6}[\hat{\eta}^{eq}_{x}-\hat{\eta}_{x}],\nonumber\\
\widehat{h}_2&=&\frac{\omega_2^\phi}{6}[\hat{\eta}^{eq}_{y}-\hat{\eta}_{y}],\nonumber\\
\widehat{h}_3&=&\frac{\omega_3^\phi}{12}[(\hat{\eta}^{eq}_{xx}+\hat{\eta}^{eq}_{yy})-(\hat{\eta}_{xx}+\hat{\eta}_{yy})]+(u_x\widehat{h}_1+u_y\widehat{h}_2),\nonumber\\
\widehat{h}_4&=&\frac{\omega_4^\phi}{4}[(\hat{\eta}^{eq}_{xx}-\hat{\eta}^{eq}_{yy})-(\hat{\eta}_{xx}-\hat{\eta}_{yy})]+3(u_x\widehat{h}_1-u_y\widehat{h}_2),\nonumber\\
\widehat{h}_5&=&\frac{\omega_5^\phi}{4}[\hat{\eta}^{eq}_{xy}-\hat{\eta}_{xy}]+\frac{3}{2}(u_x\widehat{h}_2+u_y\widehat{h}_1),\nonumber\\
\widehat{h}_6&=&-\frac{\omega_6^\phi}{4}[\hat{\eta}^{eq}_{xxy}-\hat{\eta}_{xxy}]-2u_x\widehat{h}_5-\frac{1}{2}u_y\widehat{h}_4-\frac{3}{2}u_y\widehat{h}_3\nonumber\\
             &&+(1+3u_x^2/2)\widehat{h}_2+3u_xu_y\widehat{h}_1,\nonumber\\
\widehat{h}_7&=&-\frac{\omega_7^\phi}{4}[\hat{\eta}^{eq}_{xyy}-\hat{\eta}_{xyy}]-2u_y\widehat{h}_5+\frac{1}{2}u_x\widehat{h}_4-\frac{3}{2}u_x\widehat{h}_3\nonumber\\
             &&+3u_xu_y\widehat{h}_2+(1+3u_y^2/2)\widehat{h}_1,\nonumber\\
\widehat{h}_8&=&\frac{\omega_8^\phi}{4}[\hat{\eta}^{eq}_{xxyy}-\hat{\eta}_{xxyy}]-2u_x\widehat{h}_7-2u_y\widehat{h}_6-4u_xu_y\widehat{h}_5\nonumber\\
             &&+\frac{1}{2}\left(u_x^2-u_y^2\right)\widehat{h}_4-\left(2+3(u_x^2+u_y^2)/2\right)\widehat{h}_3\nonumber\\
             &&+(2+3u_x^2)u_y\widehat{h}_2+(2+3u_y^2)u_x\widehat{h}_1,
\end{eqnarray}
where $\omega_1^\phi=\omega_2^\phi$ controls the diffusivity $D_{\phi}=c_{s\phi}^2\left(\frac{1}{\omega_j^\phi}-\frac{1}{2}\right)\delta _t$, where $j=1,2$, while the relaxation parameters for the higher order moments $\omega_3^\phi=\omega_4^\phi$, $\omega_5^\phi$, $\omega_6^\phi$, $\omega_7^\phi$ and $\omega_8^\phi$ can be adjusted to improve numerical stability. Finally, expanding $(\tensor{K}\cdot\widehat{\mathbf{h}})_\alpha$ in Eq.~(\ref{eq:cascaded_LBMa}), the updates for the post-collision distribution functions read as
\begin{eqnarray}
\tilde{g}_0&=& g_0+[\hat{h}_0-4(\hat{h}_3-\hat{h}_8)],\nonumber\\
\tilde{g}_1&=& g_1+[\hat{h}_0+\hat{h}_1-\hat{h}_3+\hat{h}_4+2(\hat{h}_7-\hat{h}_8)],\nonumber\\
\tilde{g}_2&=& g_2+[\hat{h}_0+\hat{h}_2-\hat{h}_3-\hat{h}_4+2(\hat{h}_6-\hat{h}_8)],\nonumber\\
\tilde{g}_3&=& g_3+[\hat{h}_0-\hat{h}_1-\hat{h}_3+\hat{h}_4-2(\hat{h}_7+\hat{h}_8)],\nonumber\\
\tilde{g}_4&=& g_4+[\hat{h}_0-\hat{h}_2-\hat{h}_3-\hat{h}_4-2(\hat{h}_6+\hat{h}_8)],\nonumber\\
\tilde{g}_5&=& g_5+[\hat{h}_0+\hat{h}_1+\hat{h}_2+2\hat{h}_3+\hat{h}_5-\hat{h}_6-\hat{h}_7+\hat{h}_8],\nonumber\\
\tilde{g}_6&=& g_6+[\hat{h}_0-\hat{h}_1+\hat{h}_2+2\hat{h}_3-\hat{h}_5-\hat{h}_6+\hat{h}_7+\hat{h}_8],\nonumber\\
\tilde{g}_7&=& g_7+[\hat{h}_0-\hat{h}_1-\hat{h}_2+2\hat{h}_3+\hat{h}_5+\hat{h}_6+\hat{h}_7+\hat{h}_8],\nonumber\\
\tilde{g}_8&=& g_8+[\hat{h}_0+\hat{h}_1-\hat{h}_2+2\hat{h}_3-\hat{h}_5+\hat{h}_6-\hat{h}_7+\hat{h}_8].
\end{eqnarray}

\section{\label{app:Central_Moment_LBM_CDE}Non-cascaded central moment LBM for solution of scalar transport to recover the skew-symmetric velocity gradient tensor}
For completeness, we will also present another version of a LBM based on central moments for solving the transport of the scalar field that allows
local computation of the vorticity. Unlike~\ref{app:Cascaded_LBM_CDE}, the formulation given below is non-cascaded, i.e., the change of
higher moments under collision do not depend on those of the lower moments. Rather, it is based on the relaxation of various central moments to their equilibria under collision, while involving systematic transformations between the distribution functions, raw moments and central moments before and after collision (similar to the algorithms presented in~\cite{Geier2015}). In this regard, we first enumerate the distribution functions, bare raw moments and central moments for the D2Q9 lattice, represented by vectors $\mathbf{g}$, $\hat {\mathbf {q}}$ and $\hat {\mathbf {q}}^c$, respectively, as
\begin{equation}
\mathbf{g} = (g_0,g_1,g_2,\cdots,g_8)^\dagger,\label{eq:distributionfunctionsCDE}
\end{equation}
\begin{eqnarray}
\hat {\mathbf {q}} &=&(\hat {q}_0,\hat {q}_1,\hat {q}_2\dots \hat {q}_8)^\dagger\nonumber\\
&=&(\hat{\eta}_0^\prime,\hat{\eta}_x^\prime,\hat{\eta}_y^\prime,\hat{\eta}_{xx}^\prime,\hat{\eta}_{yy}^\prime,\hat{\eta}_{xy}^\prime,\hat{\eta}_{xxy}^\prime,\hat{\eta}_{xyy}^\prime,\hat{\eta}_{xxyy}^\prime)^\dagger,\label{eq:barerawmomentdefnitionCDE}
\end{eqnarray}
\begin{eqnarray}
\hat {\mathbf {q}}^c &=&(\hat {q}_0^c,\hat {q}_1^c,\hat {q}_2^c\dots \hat {q}_8^c)^\dagger,\nonumber\\
&=&(\hat{\eta}_0,\hat{\eta}_x,\hat{\eta}_y,\hat{\eta}_{xx},\hat{\eta}_{yy},\hat{\eta}_{xy},\hat{\eta}_{xxy},\hat{\eta}_{xyy},\hat{\eta}_{xxyy})^\dagger.\label{eq:barecentralmomentdefnitionCDE}
\end{eqnarray}
Then, the mappings between the central moments, raw moments and distribution functions may be formally expressed in matrix-vector forms as
\begin{equation}
\hat {\mathbf {q}}^c = \tensor{\mathcal{F}} \hat {\mathbf {q}},\quad \hat {\mathbf {q}} = \tensor{\mathcal{F}}^{-1} \hat {\mathbf {q}}^c,\quad \hat {\mathbf {q}} = \tensor{P} \mathbf{g},\quad \mathbf {g} = \tensor{P}^{-1} \hat{\mathbf{q}},\label{eq:mappingrelationsdfmomentsCDE}
\end{equation}
where $\tensor{P}$ is a matrix representing the transformation from the distribution functions to the raw moments (see Eqs.~(\ref{eq:transformationmatrix1}) and (\ref{eq:basisvectors1})) and $\tensor{\mathcal{F}}$ is a frame transformation matrix that maps the raw moments to the central moments,  i.e., containing the elements of $(\mathbf{e_x}-u_x\mathbf{1})^m(\mathbf{e_y}-u_y\mathbf{1})^n$. However, since
$\tensor{P}$ and $\tensor{P}^{-1}$ are both sparse, while $\tensor{\mathcal{F}}$ as well as $\tensor{\mathcal{F}}^{-1}$ are of special lower triangular forms arising from the binomial expansions, it is neither necessary nor efficient to use them in matrix forms. Rather, we only list the resulting mapping expressions of the elements of each transformation before and after collision in the algorithm in what follows.

\subsection*{\underline{(a) Pre-collision raw moments}}
Expanding $\hat {\mathbf {q}} = \tensor{P} \mathbf{g}$, the raw moments before collision read as
\begin{eqnarray}
\hat{\eta}^\prime_{0} &=& g_0+g_1+g_2+g_3+g_4+s_g,\nonumber\\
\hat{\eta}^\prime_{x} &=& g_1-g_3+g_5-g_6-g_7+g_8,\nonumber\\
\hat{\eta}^\prime_{y} &=& g_2-g_4+g_5-g_6-g_7-g_8,\nonumber\\
\hat{\eta}^\prime_{xx} &=& g_1+g_3+s_g,\nonumber\\
\hat{\eta}^\prime_{yy} &=& g_2+g_4+s_g,\nonumber\\
\hat{\eta}^\prime_{xy} &=& g_5-g_6+g_7-g_8,\nonumber\\
\hat{\eta}^\prime_{xxy} &=& g_5+g_6-g_7-g_8,\nonumber\\
\hat{\eta}^\prime_{xyy} &=& g_5-g_6-g_7+g_8,\nonumber\\
\hat{\eta}^\prime_{xxyy} &=& s_g,
\end{eqnarray}
where
\begin{equation*}
s_g = g_5+g_6+g_7+g_8
\end{equation*}

\subsection*{\underline{(b) Pre-collision central moments}}
Based on $\hat {\mathbf {q}}^c = \tensor{\mathcal{F}} \hat {\mathbf {q}}$, the central moments from raw moments before collision follows. Hence, we obtain
\begin{eqnarray}
\hat{\eta}_{0} &=& \hat{\eta}^\prime_{0},\nonumber\\
\hat{\eta}_{x} &=& \hat{\eta}_{x}^\prime-u_x\hat{\eta}_{0}^\prime,\nonumber\\
\hat{\eta}_{y} &=& \hat{\eta}_{y}^\prime-u_y\hat{\eta}_{0}^\prime,\nonumber\\
\hat{\eta}_{xx} &=& \hat{\eta}_{xx}^\prime-2u_x\hat{\eta}_{x}^\prime+u_x^2\hat{\eta}_{0}^\prime,\nonumber\\
\hat{\eta}_{yy} &=& \hat{\eta}_{yy}^\prime-2u_y\hat{\eta}_{y}^\prime+u_y^2\hat{\eta}_{0}^\prime,\nonumber\\
\hat{\eta}_{xy} &=& \hat{\eta}_{xy}^\prime-u_y\hat{\eta}_{x}^\prime-u_x\hat{\eta}_{y}^\prime+u_xu_y\hat{\eta}_{0}^\prime,\nonumber\\
\hat{\eta}_{xxy} &=& \hat{\eta}_{xxy}^\prime-2u_x\hat{\eta}_{xy}^\prime+u_x^2\hat{\eta}_{y}^\prime-u_y\hat{\eta}_{xx}^\prime+2u_xu_y\hat{\eta}_{x}^\prime-u_x^2u_y\hat{\eta}_{0}^\prime,\nonumber\\
\hat{\eta}_{xyy} &=& \hat{\eta}_{xyy}^\prime-2u_y\hat{\eta}_{xy}^\prime+u_y^2\hat{\eta}_{x}^\prime-u_x\hat{\eta}_{yy}^\prime+2u_xu_y\hat{\eta}_{y}^\prime-u_xu_y^2\hat{\eta}_{0}^\prime,\nonumber\\
\hat{\eta}_{xxyy}&=&\hat{\eta}_{xxyy}^\prime-2u_x\hat{\eta}_{xyy}^\prime-2u_y\hat{\eta}_{xxy}^\prime+u_x^2\hat{\eta}_{yy}^\prime+u_y^2\hat{\eta}_{xx}^\prime\nonumber\\
                 &&+4u_xu_y\hat{\eta}_{xy}^\prime-2u_x^2u_y\hat{\eta}_{y}^\prime-2u_xu_y^2\hat{\eta}_{x}^\prime+u_x^2u_y^2\hat{\eta}_{0}^\prime
\end{eqnarray}

\subsection*{\underline{(c) Post-collision central moments: Relaxation of central moments under collision}}
We then prescribe the relaxation of various central moments to their equilibria at individual rates under collision, where the central moment equilibria that account for the anisotropy at the third order to recover the vorticity field are given in Eq.~(\ref{eq:eqmcentralmoment_CDE}). Hence, the post-collision central moments can be written as
\begin{eqnarray}
\widetilde{\hat{\eta}}_{0} &=& \hat{\eta}_{0} \nonumber\\
\widetilde{\hat{\eta}}_{x} &=& \hat{\eta}_{x} +\omega_1^\phi[\hat{\eta}_{x}^{eq}-\hat{\eta}_{x}],\nonumber\\
\widetilde{\hat{\eta}}_{y} &=& \hat{\eta}_{y} +\omega_2^\phi[\hat{\eta}_{y}^{eq}-\hat{\eta}_{y}],\nonumber\\
\widetilde{\hat{\eta}}_{xx}+\widetilde{\hat{\eta}}_{yy} &=& (\hat{\eta}_{xx}+\hat{\eta}_{yy})+\omega_3^\phi[(\hat{\eta}_{xx}^{eq}+\hat{\eta}_{yy}^{eq})-(\hat{\eta}_{xx}+\hat{\eta}_{yy})],\nonumber\\
\widetilde{\hat{\eta}}_{xx}-\widetilde{\hat{\eta}}_{yy} &=& (\hat{\eta}_{xx}-\hat{\eta}_{yy})+\omega_4^\phi[(\hat{\eta}_{xx}^{eq}-\hat{\eta}_{yy}^{eq})-(\hat{\eta}_{xx}-\hat{\eta}_{yy})],\nonumber\\
\widetilde{\hat{\eta}}_{xy} &=& \hat{\eta}_{xy} + \omega_5^\phi[\hat{\eta}_{xy}^{eq}-\hat{\eta}_{xy}],\nonumber\\
\widetilde{\hat{\eta}}_{xxy} &=& \hat{\eta}_{xxy} + \omega_6^\phi[\hat{\eta}_{xxy}^{eq}-\hat{\eta}_{xxy}],\nonumber\\
\widetilde{\hat{\eta}}_{xyy} &=& \hat{\eta}_{xyy} + \omega_7^\phi[\hat{\eta}_{xyy}^{eq}-\hat{\eta}_{xyy}],\nonumber\\
\widetilde{\hat{\eta}}_{xxyy} &=& \hat{\eta}_{xxyy} + \omega_8^\phi[\hat{\eta}_{xxyy}^{eq}-\hat{\eta}_{xxyy}].
\end{eqnarray}
The choices of the various relaxation times $\omega_j^\phi$, where $j=1,2,\ldots,8$ are the same as those given in~\ref{app:Cascaded_LBM_CDE}.

\subsection*{\underline{(d) Post-collision raw moments}}
The post-collision central moments can be mapped to those of raw moments via $\widetilde{\hat {\mathbf {q}}} = \tensor{\mathcal{F}}^{-1} \widetilde{\hat {\mathbf {q}}}^c$. It may be noted that the elements of $\tensor{\mathcal{F}}^{-1}$ (representing the inverse of binomial expansions) are the same of those of $\tensor{\mathcal{F}}$ (representing the binomial expansions)  after making all the coefficients
in the latter to be positive. Hence, we get
\begin{eqnarray}
\widetilde{\hat{\eta}}_{0}^\prime &=& \widetilde{\hat{\eta}}_{0},\nonumber\\
\widetilde{\hat{\eta}}_{x}^\prime &=& \widetilde{\hat{\eta}}_{x}+u_x\widetilde{\hat{\eta}}_{0},\nonumber\\
\widetilde{\hat{\eta}}_{y}^\prime &=& \widetilde{\hat{\eta}}_{y}+u_y\widetilde{\hat{\eta}}_{0},\nonumber\\
\widetilde{\hat{\eta}}_{xx}^\prime &=& \widetilde{\hat{\eta}}_{xx}+2u_x\widetilde{\hat{\eta}}_{x}+u_x^2\widetilde{\hat{\eta}}_{0},\nonumber\\
\widetilde{\hat{\eta}}_{yy}^\prime &=& \widetilde{\hat{\eta}}_{yy}+2u_y\widetilde{\hat{\eta}}_{y}+u_y^2\widetilde{\hat{\eta}}_{0},\nonumber\\
\widetilde{\hat{\eta}}_{xy}^\prime &=& \widetilde{\hat{\eta}}_{xy}+u_y\widetilde{\hat{\eta}}_{x}+u_x\widetilde{\hat{\eta}}_{y}+u_xu_y\widetilde{\hat{\eta}}_{0},\nonumber\\
\widetilde{\hat{\eta}}_{xxy}^\prime &=& \widetilde{\hat{\eta}}_{xxy}+2u_x\widetilde{\hat{\eta}}_{xy}+u_x^2\widetilde{\hat{\eta}}_{y}+u_y\widetilde{\hat{\eta}}_{xx}+2u_xu_y\widetilde{\hat{\eta}}_{x}+u_x^2u_y\widetilde{\hat{\eta}}_{0},\nonumber\\
\widetilde{\hat{\eta}}_{xyy}^\prime &=& \widetilde{\hat{\eta}}_{xyy}+2u_y\widetilde{\hat{\eta}}_{xy}+u_y^2\widetilde{\hat{\eta}}_{x}+u_x\widetilde{\hat{\eta}}_{yy}+2u_xu_y\widetilde{\hat{\eta}}_{y}+u_xu_y^2\widetilde{\hat{\eta}}_{0},\nonumber\\
\widetilde{\hat{\eta}}_{xxyy}^\prime&=&\widetilde{\hat{\eta}}_{xxyy}+2u_x\widetilde{\hat{\eta}}_{xyy}+2u_y\widetilde{\hat{\eta}}_{xxy}+u_x^2\widetilde{\hat{\eta}}_{yy}+u_y^2\widetilde{\hat{\eta}}_{xx}\nonumber\\
                 &&+4u_xu_y\widetilde{\hat{\eta}}_{xy}+2u_x^2u_y\widetilde{\hat{\eta}}_{y}+2u_xu_y^2\widetilde{\hat{\eta}}_{x}+u_x^2u_y^2\widetilde{\hat{\eta}}_{0}
\end{eqnarray}

\subsection*{\underline{(e) Post-collision distribution functions}}
Finally, the post-collision distribution functions can be obtained by simplifying $\mathbf {\widetilde{g}} = \tensor{P}^{-1} \widetilde{\hat{\mathbf{q}}}$, which yield
\begin{eqnarray}
\widetilde{g}_0 &=& \widetilde{\hat{\eta}}_{0}^\prime-\widetilde{\hat{\eta}}_{xx}^\prime-\widetilde{\hat{\eta}}_{yy}^\prime+\widetilde{\hat{\eta}}_{xxyy}^\prime,\nonumber\\
\widetilde{g}_1 &=& \frac{1}{2}\left( \widetilde{\hat{\eta}}_{x}^\prime+\widetilde{\hat{\eta}}_{xx}^\prime-\widetilde{\hat{\eta}}_{xyy}^\prime-\widetilde{\hat{\eta}}_{xxyy}^\prime \right),\nonumber\\
\widetilde{g}_2 &=& \frac{1}{2}\left( \widetilde{\hat{\eta}}_{y}^\prime+\widetilde{\hat{\eta}}_{yy}^\prime-\widetilde{\hat{\eta}}_{xxy}^\prime-\widetilde{\hat{\eta}}_{xxyy}^\prime \right),\nonumber\\
\widetilde{g}_3 &=& \frac{1}{2}\left( -\widetilde{\hat{\eta}}_{x}^\prime+\widetilde{\hat{\eta}}_{xx}^\prime+\widetilde{\hat{\eta}}_{xyy}^\prime-\widetilde{\hat{\eta}}_{xxyy}^\prime \right),\nonumber\\
\widetilde{g}_4 &=& \frac{1}{2}\left( -\widetilde{\hat{\eta}}_{y}^\prime+\widetilde{\hat{\eta}}_{yy}^\prime+\widetilde{\hat{\eta}}_{xxy}^\prime-\widetilde{\hat{\eta}}_{xxyy}^\prime \right),\nonumber\\
\widetilde{g}_5 &=& \frac{1}{4}\left( \widetilde{\hat{\eta}}_{xy}^\prime+\widetilde{\hat{\eta}}_{xxy}^\prime+\widetilde{\hat{\eta}}_{xyy}^\prime+\widetilde{\hat{\eta}}_{xxyy}^\prime \right),\nonumber\\
\widetilde{g}_6 &=& \frac{1}{4}\left( -\widetilde{\hat{\eta}}_{xy}^\prime+\widetilde{\hat{\eta}}_{xxy}^\prime-\widetilde{\hat{\eta}}_{xyy}^\prime+\widetilde{\hat{\eta}}_{xxyy}^\prime \right),\nonumber\\
\widetilde{g}_7 &=& \frac{1}{4}\left( \widetilde{\hat{\eta}}_{xy}^\prime-\widetilde{\hat{\eta}}_{xxy}^\prime-\widetilde{\hat{\eta}}_{xyy}^\prime+\widetilde{\hat{\eta}}_{xxyy}^\prime \right),\nonumber\\
\widetilde{g}_8 &=& \frac{1}{4}\left( -\widetilde{\hat{\eta}}_{xy}^\prime-\widetilde{\hat{\eta}}_{xxy}^\prime+\widetilde{\hat{\eta}}_{xyy}^\prime+\widetilde{\hat{\eta}}_{xxyy}^\prime \right).
\end{eqnarray}

\end{document}